# Longitudinal and transverse electron paramagnetic resonance in a scanning tunneling microscope

Tom S. Seifert,[1]* Stepan Kovarik,[1] Dominik M. Juraschek,[1,2] Nicola A. Spaldin,[1] Pietro Gambardella,[1] Sebastian Stepanow[1]*

[1] Department of Materials, ETH Zürich, 8093 Zürich, Switzerland

[2] Harvard John A. Paulson School of Engineering and Applied Sciences, Harvard University, Cambridge, Massachusetts 02138, USA

*Corresponding author. Email: tom.seifert@mat.ethz.ch; sebastian.stepanow@mat.ethz.ch

**Electron paramagnetic resonance (EPR) spectroscopy is widely employed to characterize paramagnetic complexes. Recently, EPR combined with scanning tunneling microscopy (STM) achieved single-spin sensitivity with sub-angstrom spatial resolution. The excitation mechanism of EPR in STM, however, is broadly debated, raising concerns about widespread application of this technique. Here, we present an extensive experimental study and modelling of EPR-STM of Fe and hydrogenated Ti atoms on an MgO surface. Our results support a piezoelectric coupling mechanism, in which the EPR species oscillate adiabatically in the inhomogeneous magnetic field of the STM tip. An analysis based on Bloch equations combined with atomic-multiplet calculations identifies different EPR driving forces. Specifically, transverse magnetic-field gradients drive the spin-1/2 hydrogenated Ti, whereas longitudinal magnetic-field gradients drive the spin-2 Fe. Additionally, our results highlight the potential of piezoelectric coupling to induce electric dipole moments, thereby broadening the scope of EPR-STM to nonpolar species and nonlinear excitation schemes.**

## INTRODUCTION

Combining the nanometer spatial resolution of a scanning tunneling microscope (STM) with the outstanding energy resolution of electron paramagnetic resonance (EPR) allows for the study of magnetic properties and interactions at the atomic scale with sensitivity to excitations surpassing the thermal resolution limit of STM by orders of magnitude (*1*). EPR-STM has been successfully used to study transition-metal atoms adsorbed on MgO and their interactions (*1-3*) using a resonant continuous-wave radio frequency (rf) excitation (*4-11*). Moreover, pulsed rf schemes have been used to coherently drive EPR excitations in single atoms (*12*). These developments open the way for further applications of EPR-STM, including the storage and retrieval of quantum information from surface spins (*13*), measurements of the relaxation time of single molecule magnets (*14, 15*), and the characterization of active sites and intermediate reaction species in catalysis (*16, 17*).

Despite early EPR-STM proposals employing nonmagnetic tips (*18, 19*) and recent experimental achievements using spin-polarized tips (*1-12*), the driving mechanism of EPR-STM remains under debate. The central idea of EPR is that radiofrequency (rf) photons excite unpaired electrons to a higher energy spin state, which can be probed experimentally. The mechanisms underpinning the excitation and detection of EPR within an STM junction under simultaneous direct current (dc) and rf bias, however, are not directly evident, particularly because the direct excitation of the EPR species by the magnetic field components of the rf tunneling and displacement currents are estimated to be negligible (*7, 10*). In addition, the scattering of tunneling electrons at the spin center is relatively strong, thus disturbing the free evolution of the magnetic states. Reproducible EPR-STM experiments require the use of a magnetic tip (*1-12*), which further complicates the modelling of the STM junction. Several EPR-STM excitation and detection mechanisms have been proposed (*19-26*), including modulation of the tunneling barrier by the rf electric field (*23*), breathing of the density of states mediated by spin-orbit coupling (*19*), spin torque due to tunneling electrons (*24*), and piezoelectric coupling (PEC) of the rf electric field to the magnetic adatom (*22*). In the PEC mechanism, the oscillating electric field couples to the electric dipole of the EPR species and induces vibrations in the inhomogeneous magnetic field of the nearby magnetic STM tip leading to an effective oscillating magnetic field that drives the EPR transitions. In the original work in Ref. (*1*), the electric-field induced motion of the atoms was already proposed, however, the EPR transitions were ascribed to modulations of the crystal field operators. Supporting experimental data for each of these mechanisms are limited. Yang *et al.* (*7*) analysed EPR-STM spectra of hydrogenated Ti based on the PEC model and found a disagreement by a factor of 40 with the rf atomic displacement calculated by theory within the harmonic approximation of the local bond vibrations. Also, Willke *et al.* (*8*) concluded that the EPR-STM driving force for hydrogenated Ti is proportional to the tunneling current, which is consistent with the PEC model (*4, 22*), but the limited experimental data did not allow for a conclusive proof and discrimination from other models. More importantly, it is still an open question which selection rules apply for the EPR-STM transitions, e.g., between the high spin and orbital moment ground state doublet of Fe on MgO/Ag(100) as compared to conventional EPR where only magnetic-dipole allowed transitions are accessible (*1*). These shortcomings, and the importance of designing future EPR-STM investigations based on the correct model, call for a comprehensive experimental and modelling approach to explore the full parameter space of EPR-STM to reveal the driving mechanism.

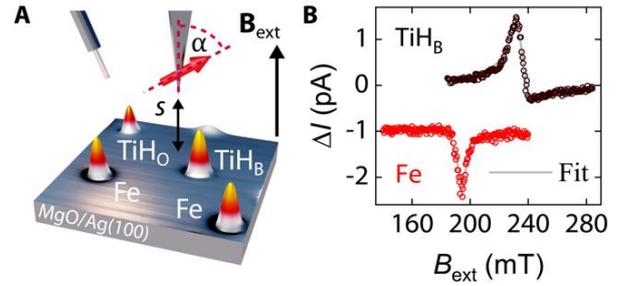

**Fig. 1. Principle of EPR-STM and representative EPR spectra.** (**A**) Schematic of the STM junction showing single magnetic adatoms on a double-layer MgO on Ag(100) driven by an rf antenna using a spin-polarized tip. The tip is at a standoff distance $s$ from point contact with the surface. The tip magnetization makes an angle α with respect to the out-of-plane external magnetic field $B_{\text{ext}}$. The schematic includes a three-dimensional rendering of a constant-current STM image (10x10 nm$^2$) of Fe and hydrogenated Ti (TiH) adatoms, on which all EPR measurements are performed (subscripts O and B label apical and bridge binding sites relative to the oxygen lattice, respectively). Setpoint current 50 pA, dc bias 30 mV. (**B**) Representative EPR spectra measured by sweeping $B_{\text{ext}}$ on TiH$_B$ and Fe adatoms shown in (A) (left Fe adatom) while applying an rf voltage to the antenna. The solid lines are fits to the data (see text). A nonresonant background is subtracted from both spectra; for clarity, the Fe spectrum is offset by −1 pA. Settings: $I_{\text{dc}} = 70$ pA, $V_{\text{dc}} = 160$ mV, $V_{\text{rf}} = 256$ mV, $\omega_{\text{rf}}/2\pi$: 8 GHz for Ti$_B$, 36 GHz for Fe.

In this work, we present a combined experimental and theoretical EPR-STM investigation of single Fe and hydrogenated Ti (TiH) adatoms adsorbed on 2 monolayers of MgO/Ag(100). To limit the number of free parameters, we perform the measurements using the same magnetic tip and employ a broad range of excitation conditions, which allows us to identify the dominant EPR excitation sources. An extended discussion of the role of the magnetic tip is reported in Section S1 of the Supplementary Material. We choose Fe and Ti since they have been characterized previously by EPR-STM (*1, 2, 10*). Fe atoms on MgO/Ag(100) were also studied by x-ray magnetic circular dichroism, inelastic tunneling spectroscopy, and ligand field theory, which provide a consistent description of the Fe wave functions within an atomic-multiplet model (*27*). In contrast to most previous EPR-STM works, we employ an external magnetic field that is strictly out of plane. To assess our data against different EPR-STM mechanisms, we completely characterize the vector magnetic field of the STM tip, including exchange and dipolar contributions, and extract the Rabi frequency $\Omega$ from the EPR spectra, thus inferring the EPR driving force for our broad range of experimental conditions. This includes an extensive analysis of the dependence of the EPR signal on the external magnetic field $B_{\text{ext}}$, the rf-voltage amplitude $V_{\text{rf}}$, the dc voltage $V_{\text{dc}}$, and the dc setpoint current $I_{\text{dc}}$. Thereby we acquired more than 100 spectra within the EPR-STM parameter space using the same magnetic microtip at different standoff distances ($s$) from the surface, where $s$ is the distance from point contact between the adatom and the STM tip. A qualitative assessment of different EPR-STM mechanisms shows that the Rabi frequency is consistent with a PEC model. To provide a more stringent quantitative comparison between theory and experiment, we evaluate the Rabi rate predicted by the PEC model by computing the rf-electric-field-induced displacement of the EPR species from first principles without relying on the harmonic approximation of the phonon dispersion curves and determining the relevant EPR transition matrix elements by multiplet calculations. We find quantitative agreement with our entire data set. Importantly, our theoretical treatment reveals that state mixing enables EPR transitions between magnetic-dipole-forbidden states as in Fe. In such a spin $S = 2$ system, the longitudinal tip-magnetic-field gradient drives EPR, in contrast to $S = 1/2$ systems such as TiH, where the transverse tip-magnetic-field gradient causes EPR excitations. Our theory also predicts nonlinear driving forces through coupling to induced electric dipoles, which potentially opens this technique to the investigation of nonpolar systems.

## RESULTS

### Recording EPR spectra with an STM

Our EPR-STM setup is depicted in Fig. 1A: A spin-polarized tip is positioned above a magnetic adatom adsorbed on a double layer of MgO on Ag(100) (*10*). A magnetic field splits the atomic energy levels by the Zeeman interaction and a resonant rf excitation induces transitions between these split states. EPR spectra are acquired by sweeping the out-of-plane $\mathbf{B}_{\text{ext}}$ while keeping the rf frequency $\omega_{\text{rf}}/2\pi$ constant and detecting the rf-induced change in the dc tunneling current $\Delta I$ using a modulation scheme of the rf source followed by a lock-in detection (*10*). Figure 1B shows typical constant-frequency EPR spectra on Fe and TiH. Following Ref. (*4*), EPR is detected electrically through the tunneling magnetoresistance (TMR) of the tip-adatom junction. Since the conductance of the STM junction depends on the relative alignment of the magnetic moments of the tip and adatom, the EPR dynamics induces a change of both the dc and ac TMR. The dc TMR variation is caused by the time-averaged population change of the magnetic-adatom states and is detected as a change of the dc tunneling current. The ac TMR originates from the rf conductance change and gives rise to an additional (homodyne) dc tunneling current via mixing with the rf voltage. For further experimental details see Materials and Methods and Ref. (*10*).

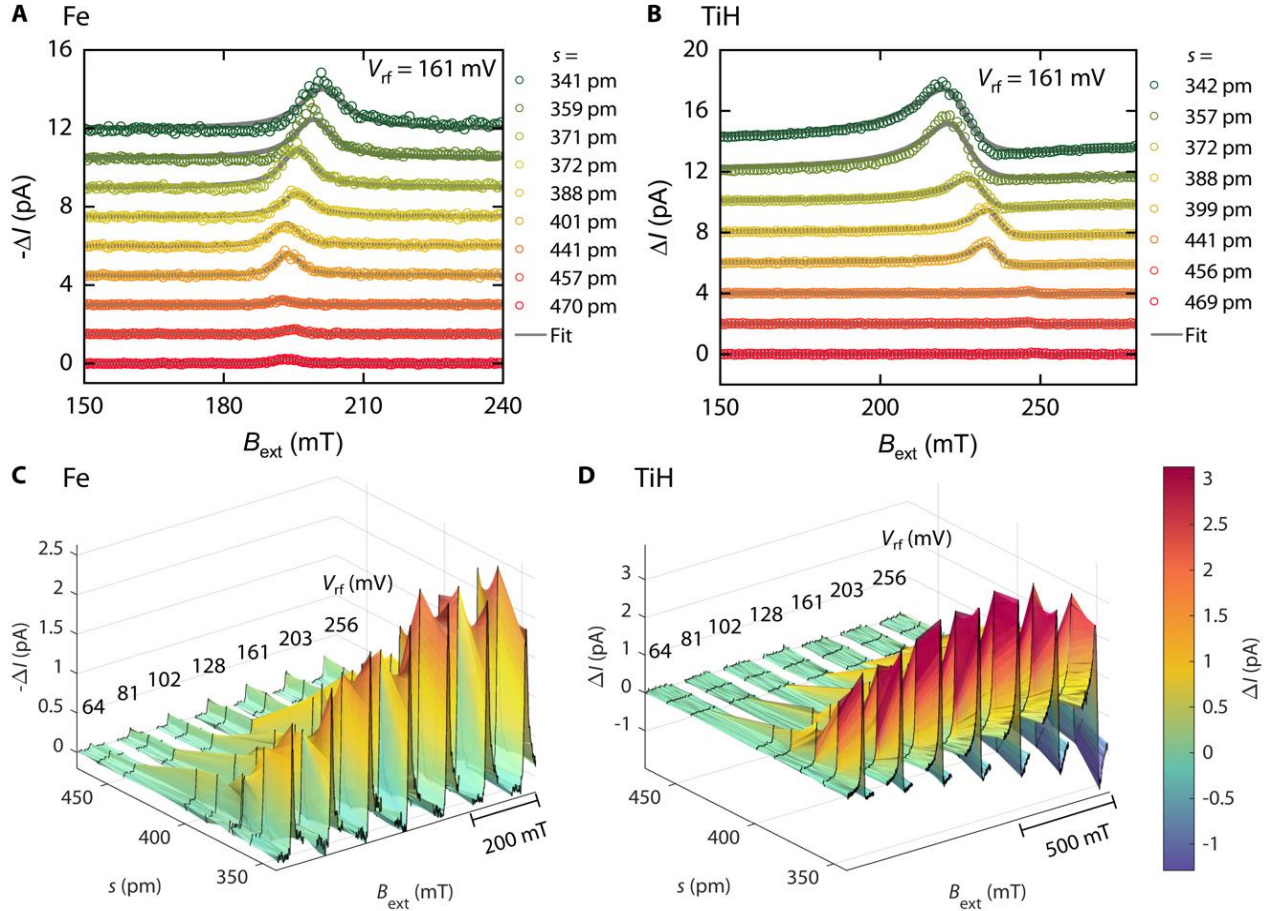

**Fig. 2. EPR data set on Fe and hydrogenated Ti.** EPR spectra of Fe at $\omega_{rf}/2\pi = 36$ GHz (**A, C**) and TiH at 8 GHz (**B, D**) recorded with the same microtip for varying the standoff distance $s$ and rf-voltage amplitude $V_{rf}$. For better visibility, the Fe spectra are inverted. **A-B** show data for a constant rf-voltage amplitude of $V_{rf} = 161$ mV. The spectra are vertically offset for better visibility. In **C-D**, the spectra are offset along the $B_{ext}$-axis for distinct values of $V_{rf}$. Rows from left to right correspond to $V_{rf} = 64$ mV, 81 mV, 102 mV, 128 mV, 161 mV, 203 mV and 256 mV.

To discriminate between different EPR mechanisms, we first have to characterize the basic elements of an EPR-STM experiment, i.e., the rf excitation, the EPR species, and the magnetic tip. The rf excitation is provided by an antenna capacitively coupled to the STM tip and is well-understood from a previous study (*10*). Also, Fe and TiH adatoms are two well-known, yet magnetically-distinct systems (*1, 2, 10*). However, the structure of the magnetic tip is completely unknown and requires further characterization. To this end, we record an extensive EPR data set on the Fe and TiH adatoms as shown in Fig. 1A using the same magnetic microtip at a similar external magnetic field, resulting in 119 spectra in the EPR-STM parameter space (see Fig. 2). Note that spectra for different $V_{rf}$ are offset for better visibility in Fig. 2 (C-D). Without any further analysis, these spectra already reveal important characteristic features of EPR-STM: (i) The amplitude and width of the EPR signal of both magnetic adatoms grow with increasing rf-voltage amplitude $V_{rf}$ and decreasing standoff distance $s$, indicating that the excitation is stronger close to the tip. (ii) The external magnetic field at resonance changes by less than 20 mT with either $s$ or $V_{rf}$, ruling out tip and bias-induced changes of the electronic ground state of the probed magnetic adatoms. (iii) The peak-to-peak amplitude is about twice as large for TiH as for Fe. (iv) The EPR spectra of TiH have opposite sign than those of Fe (note that the Fe spectra are inverted in Fig. 2A and C) and (v) the EPR signal line shape of TiH has a stronger dependence on *s* than that of Fe. (vi) The EPR signal is mostly symmetric for Fe and more asymmetric for TiH, for which the asymmetry grows with increasing $V_{rf}$. The same general features are observed when changing the STM tip, for all the 6 different EPR-active microtips investigated in this study (see Fig. S9). These observations indicate a different nature of EPR-STM for Fe and TiH, requiring a more detailed analysis of the recorded spectra to allow for an assessment of the EPR driving forces.

**Analysis of the EPR spectra and Rabi rate**

For a quantitative analysis of the mechanisms that drive the EPR of Fe and TiH, we fit the spectra in Fig. 2 using a general model of the change in tunneling current flowing between a magnetic adatom and a spin-polarized tip in the presence of an rf bias. According to Ref. (*4*) and as summarized in Section S2, the total rf-induced current is given by

$$\Delta I = I_{\text{off}} - a_{\text{TMR}} I_{\text{dc}} \frac{\Omega^2 T_1 T_2}{1 + \Delta\omega^2 T_2^2 + \Omega^2 T_1 T_2} \cdot$$

$$\left( \cos\alpha + \frac{\Delta\omega T_2 V_{\text{rf}}}{2\Omega T_1 V_{\text{dc}}} \sin\alpha \right), \quad (1)$$

where the first term is an offset that accounts for magnetic-field-independent rectified rf currents due to STM-junction-conductance nonlinearities. The second term describes the TMR of the STM junction modulated by spin precession. It includes a term proportional to the time-averaged projection of the atomic magnetic moment on the tip magnetization ($\sim \cos\alpha$) and a homodyne contribution ($\sim \sin\alpha$), where $\alpha$ is the angle between the tip magnetization and the external magnetic field (see Fig. 1A). $a_{\text{TMR}}$ is a parameter that describes the TMR amplitude and $I_{\text{dc}}$ is the dc setpoint current. Note that, for a vanishing $V_{\text{dc}}$, the ratio $I_{\text{dc}}/V_{\text{dc}}$ that appears in the second term approaches the dc setpoint conductance of the tunneling junction. $T_1$ and $T_2$ are the longitudinal and transverse spin lifetimes, respectively, and $\Delta\omega = 2\mu(B_{\text{ext}}^0 - B_{\text{ext}})/\hbar$ is the detuning from the external magnetic field at which the resonance occurs $B_{\text{ext}}^0$, with $\hbar$ the reduced Planck constant and $\mu$ the adatom magnetic moment. The latter is 1 $\mu_B$ for TiH (*10*) and 5.2 $\mu_B$ for Fe (*27*).

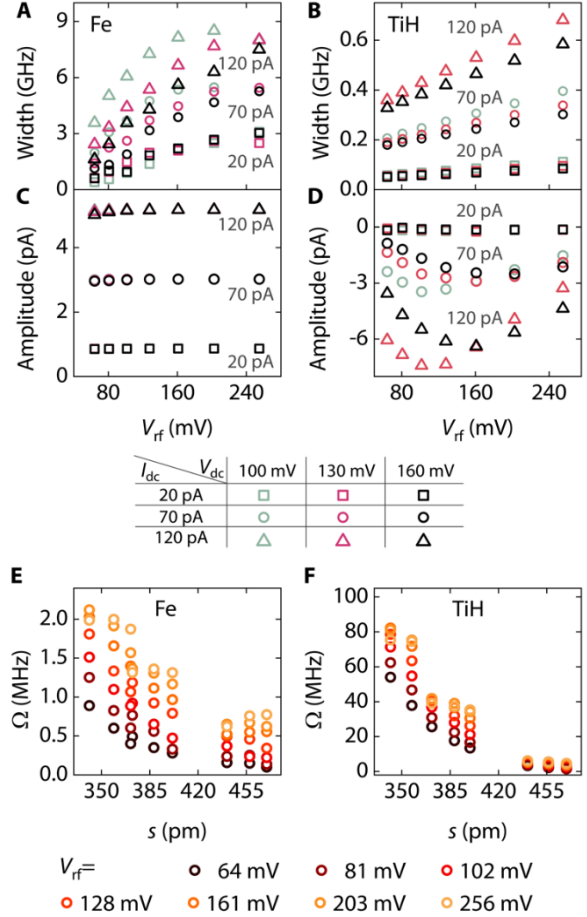

**Fig. 3. Amplitude and width of the EPR spectra and experimental Rabi rates.** A fit of all 119 EPR spectra using Eq. 1 (see text, Section S3 and Fig. S4) allows for calculating the spectral line widths (**A** and **B**) and amplitudes (**C** and **D**) for varying rf-voltage amplitude $V_{\text{rf}}$, dc voltage $V_{\text{dc}}$ and setpoint current $I_{\text{dc}}$. In C, most symbols for a given $I_{\text{dc}}$ overlap. Experimental Rabi rate $\Omega$ vs standoff distance *s* for Fe (**E**) and TiH (**F**) at different $V_{\text{rf}}$. The errors in **E-F** of $\pm 2$ % are smaller than the size of the symbols.

Equation 1 was initially derived for a $S = 1/2$ system such as TiH, but we show below that it is also valid for higher spin systems such as Fe if the Rabi frequency Ω of an effective two-level system is appropriately renormalized by the matrix element connecting the true initial and final magnetic states, as outlined in Section S2. We note also that Eq. 1 neglects a possible spin torque initialization of the magnetic-adatom spin (*5*), which would alter the EPR signal through a change of the off-resonant magnetic state population induced by inelastic spin-flip excitations by the tunneling electrons. The impact of this effect is minimized by measuring the EPR in a relatively narrow range of the dc bias voltage with constant polarity and using rf-voltage amplitudes large compared to the inelastic spin-flip thresholds. Finally, our model neglects the hyperfine interaction (*11*) which is justified since the investigated adatoms did not show an associated broadening or splitting of the EPR lines.

Given that all the spectra in Fig. 2 were acquired with the same tip, we perform a simultaneous fit of the entire set of EPR spectra based on the following assumptions: The magnetic moments of the probed adatoms are supposed to point on average along the out-of-plane external magnetic field $\mathbf{B}_{\text{ext}}$ (see Fig. 1A), which is justified for Fe owing to its large out-of-plane anisotropy (*27*) and for the isotropic TiH moment if $\mathbf{B}_{\text{ext}}$ is dominant with respect to in-plane components of the tip-induced magnetic field (*2*). We assume that the tunneling electrons are the main source of $T_1$ and $T_2$ events due to the large values of the dc and the rf current (see Fig. S3A) as also previously observed (*5, 28*). Thus, we set $T_{1,2} = e/(r_{1,2}I)$, where $r_{1,2}$ is the probability that a single tunneling electron induces a $T_{1,2}$ event and $I$ is the total current given by the sum of the dc and the root-mean-square rf current $I_{\text{rf}}$, which is obtained via the dc STM-junction conductance (see Materials and Methods). We further account for a fixed increase in EPR spectral line width through a convolution of the EPR spectra with a 4-mT-broad Gaussian. This broadening is caused by the atom-tracking scheme, in which the tip circles atop the magnetic adatom (3 mT, as deduced from typical magnetic-field gradients) and by the finite $B_{\text{ext}}$ sweep rate (1 mT) (*10*). With these assumptions, we fit all the EPR spectra with Eq. 1 using an adatom-independent value of α, adatom-specific parameters $a_{\text{TMR}}^{\text{Fe,Ti}}$, $r_1^{\text{Fe,Ti}}$ and $r_2^{\text{Fe,Ti}}$, and adatom-specific local parameters $B_0^{\text{Fe,Ti}}$, $I_{\text{off}}^{\text{Fe,Ti}}$ and $\Omega^{\text{Fe,Ti}}$ that depend additionally on $I_{\text{dc}}$, $V_{\text{dc}}$ and $V_{\text{rf}}$ (see Materials and Methods for further details). The best fit of all the 119 EPR spectra is found for α = (64 ± 2)°, $a_{\text{TMR}}^{\text{Fe}} = 0.043_{-0.004}^{+0.003}$, $r_1^{\text{Fe}} = 6_{-1}^{+2} \cdot 10^{-9}$, $r_2^{\text{Fe}} = 0.99_{-0.24}^{+0.33}$, $a_{\text{TMR}}^{\text{Ti}} = -0.70_{-0.05}^{+0.04}$, $r_1^{\text{Ti}} = 0.032_{-0.003}^{+0.003}$, $r_2^{\text{Ti}} = 1.00_{-0.04}^{+0.21}$, see Fig. 1B, and Fig. S4 and Section S3 in the Supplementary Material. Thus, for the relaxation times, we find that nearly every tunneling electron induces a $T_2$ event, whereas only a small fraction of them leads to a $T_1$ relaxation, in agreement with previous reports for Fe (*5*). For TiH, on the other hand, a difference in $T_1$ and $T_2$ can arise from the different probabilities for inelastic and elastic scattering events of the spin-polarized tunneling electrons with the adatom's spin. Note, that our model neglects relaxation mediated by phonons which are not significant due to the relatively high tunneling currents and the thin MgO support (*28*). The opposite sign of $a_{\text{TMR}}$ for Fe and TiH is a consequence of the opposite polarities observed in the raw data (Figs. 1B and 2).

From the above fit parameters, we derive three important quantities, namely the line width $\sqrt{1 + \Omega^2 T_1 T_2}/(\pi T_2)$ (Fig. 3, A and B), the spectral amplitude $a_{\text{TMR}} I_{\text{dc}} \Omega^2 T_1 T_2/(1 + \Omega^2 T_1 T_2)$ (Fig. 3, C and D), and the asymmetry $T_2 V_{\rm rf}/(2\Omega T_1 V_{\rm dc})$ (Fig. S6A). We observe that the line width grows almost linearly with the rf-voltage amplitude $V_{\rm rf}$ at constant $I_{\rm dc}$, which is a consequence of being in the strong-driving regime, i.e., $\Omega^2 T_1 T_2 \gtrsim 1$. This is consistent with the saturated amplitude for Fe for all $V_{\rm rf}$ and with the saturating amplitudes for TiH at the two lowest values of $I_{\rm dc}$ (see Fig. 3, C and D). For TiH and the highest value of $I_{\rm dc} = 120$ pA, that is for the smallest standoff distance $s$, we do not observe saturation of the amplitude at large $V_{\rm rf}$. This finding might indicate a change of the TiH-magnetization orientation due to an increased magnitude of the in-plane tip field that is not included in our analysis. The asymmetry of the EPR signal of TiH (Fig. S6B) grows linearly with $V_{\rm rf}$ and strongly depends on $I_{\rm dc}$ reflecting the intricate dependence of the Rabi rate on $I_{\rm dc}$ discussed below. Fe spectra show nearly symmetrical line shapes and accordingly have vanishing asymmetries (Fig. S6A), which is consistent with previous studies (*5*) and can be understood by the long $T_1$, i.e., small $r_1$ of Fe compared to TiH. In essence, this difference arises because a tunneling electron can induce a direct transition in TiH, which corresponds to a spin excitation with $\Delta S = 1$, whereas Fe has a large spin and orbital moment that cannot be directly excited by a single electron. The long $T_1$ of Fe suppresses the asymmetric EPR line shape originating from the homodyne component of Eq. 1. The shorter $T_1$ of TiH, on the other hand, gives an asymmetric line shape as also reported previously (*2*). Finally, the experimental Rabi rate $\Omega$ is given in Fig. 3 (E and F) and ranges from about 100 MHz for TiH to about 1 MHz for Fe, consistent with the literature (*12*). This information allows us to perform a qualitative assessment of the different proposed EPR-STM mechanisms, as described below.

**Assessment of different EPR-STM mechanisms**

We now contrast the observations summarized in Figs. 2 and 3 with the expectations for different excitation models of EPR-STM.

(i) A Rabi rate $\Omega$ induced by the ac magnetic field originating from the rf tunneling current and the rf displacement current has been discarded previously by estimating the respective magnitudes (*7, 10*). Additionally, we note that both contributions should not depend strongly on the standoff distance $s$, contrary to our measurements (see Fig. 2). Moreover, the rf magnetic field caused by the displacement current should depend monotonically on $s$, unlike what we observe for Fe in Fig. 3E. Also, the displacement current should be proportional to the frequency $\omega_{\rm rf}$, which is not consistent with EPR measurements performed at different $\omega_{\rm rf}$.

(ii) A spin-torque-mediated EPR (*24*) is expected to be proportional to $I_{\rm rf}$ and independent of $s$. Such a mechanism is unlikely given the strong dependence of $\Omega$ on $s$ at constant dc setpoint current (see Fig. S6, E and F).

(iii) A purely rf-electric-field driven EPR-STM, in which rf-induced spin-polarized tunneling electrons couple via the exchange interaction to the adatom magnetic moment, has been proposed in Ref. (*20*). This coupling can be understood as a current-induced effective magnetic field driving the EPR. However, this mechanism can be discarded since it fails to explain EPR in half-integer spin systems such as TiH.

(iv) A modulation of the density of states by the precessing spin of the magnetic adatom mediated by spin-orbit coupling (*19*) can be ruled out since it should be observable even with a nonmagnetic tip. This is not observed experimentally and is inconsistent with the results presented in Fig. 5 (A and B), which show that the resonance field depends on the distance between the magnetic tip and the EPR species.

(v) A change of the crystal field caused by adatom vibrations induced by the rf electric field (*1, 22*) should yield a Rabi rate that depends monotonically on $s$, unlike what is observed for Fe in Fig. 3E. Moreover, our multiplet calculations (see below, Materials and Methods, Section S6 and Fig. S8) indicate that the crystal-field operators yield vanishing EPR driving forces for Fe. Nevertheless, rf-induced variations of the crystal field could yield minor contributions to the Rabi rate in the case of TiH.

(vi) A modulation of the $g$-factor anisotropy of the EPR species by the vibrations induced by the rf electric field should lead to a Rabi rate that depends monotonically on $s$ since the driving electric field is proportional to $1/s$ in a simple plate-capacitor model (*25*). This is in contrast with our experimental findings for $\Omega$ shown in Fig. 3E.

(vii) In the PEC model (*22*), $\Omega$ is expected to be proportional to the conductance of the STM-junction if the adatom-tip interaction is dominated by the exchange interaction (*8*). This prediction is partly inconsistent with our experimental $\Omega$ (see Fig. S6, C and D), which might indicate an additional tip-adatom interaction such as dipolar coupling (see below). Apart from that, the PEC mechanism implies complex dependencies of $\Omega$ on the experimental parameters $I_{dc}$, $V_{dc}$, $s$ and $V_{rf}$ (*7*) that require a quantitative evaluation.

(viii) A cotunneling mechanism (*23*) as well as an open-quantum-system approach (*26*) have been proposed to describe the excitation and detection of EPR, respectively. Testing these approaches requires a detailed knowledge of the wave functions of the tip and EPR species that is experimentally difficult to obtain. However, as we will discuss later, these approaches represent more general descriptions that include some of the other mechanisms.

Based on this analysis, we conclude that mechanisms (i-vi) are not compatible with our experimental data set. Further evaluation of (vii) and of EPR-STM in general requires quantitative knowledge of the involved transition matrix elements and of the total magnetic field acting on the EPR species. We focus here on the most relevant magnetic-moment operator mediating EPR (see below) but discuss further operators in Section S6. Our analysis goes beyond an ideal $S = 1/2$ system since EPR encompasses a much larger variety of magnetic complexes with $S > 1/2$. It is thus important to determine what drives the EPR of Fe on MgO, which is known to have $S = 2$ (*27*), in order to reach a comprehensive understanding of EPR-STM.

**Transition matrix elements of EPR-STM for atoms with $S > 1/2$**

To drive EPR, we consider a perturbative oscillating magnetic field $\mathbf{B}_1$ acting on the magnetic adatom. The $\mathbf{B}_1$ field interacts with the magnetic moment of the adatom $\hat{\boldsymbol{\mu}} = -\mu_B(\hat{\mathbf{L}} + 2\hat{\mathbf{S}})/\hbar$ via the Zeeman interaction, and the corresponding interaction Hamiltonian reads $H' = \mu_B(\hat{\mathbf{L}} + 2\hat{\mathbf{S}}) \cdot \mathbf{B}_1/\hbar$, where $\hat{\mathbf{S}}$ and $\hat{\mathbf{L}}$ are the spin and orbital angular momentum operators, respectively. In the derivation of Eq. 1 for single TiH adatoms, $\mathbf{B}_1$ was assumed to be transverse to the static magnetic field $\mathbf{B}_0$ inducing the Zeeman splitting of the adatom's states, as in the standard two-level model of EPR [Ref. (*4*) and Section S2]. This assumption, however, has not been tested in detail and Ref. (*4*) makes no predictions about the requirements on $\mathbf{B}_1$ to drive EPR in Fe. For TiH on the bridge binding site (see Fig. 1A), we assume a nearly perfect physical $S = 1/2$ system due to the low binding-site symmetry. Accordingly, the two lowest states $|0\rangle$ and $|1\rangle$ are the $LS$-basis states $|M_L = 0, M_S = \pm 1/2\rangle$ with quenched orbital moment, as reported previously (2, 10). Those states are the natural eigenstates of the $\hat{L}$ and $\hat{S}$ operators and the interaction Hamiltonian becomes $H' = \frac{2\mu_B}{\hbar}\left(\hat{S}_x B_{1,x} + \hat{S}_y B_{1,y} + \hat{S}_z B_{1,z}\right)$. Since $\hat{S}_z$ is diagonal in the $|M_L = 0, M_S = \pm 1/2\rangle$ basis, it is evident that only transverse $\mathbf{B}_1$ fields can drive a transition between $|0\rangle$ and $|1\rangle$. For instance, for a transverse $\mathbf{B}_1$ field along $x$, the off-diagonal matrix element that drives the transition is $H'_{01} = \mu_B B_{1,x}(t)$ since $\hat{S}_x = \frac{\hbar}{2}\hat{\sigma}_x$ with $\hat{\sigma}_x$ being the $x$ component of the vector of Pauli matrices $\hat{\boldsymbol{\sigma}}$. Note that the transverse field oscillates in time proportionally to $\cos\omega_{rf}t$, i.e., $B_{1,x}(t) = B_{1,x}\cos\omega_{rf}t$, and that its amplitude relates to the Rabi rate according to $\hbar\Omega = \mu_B B_{1,x}$.

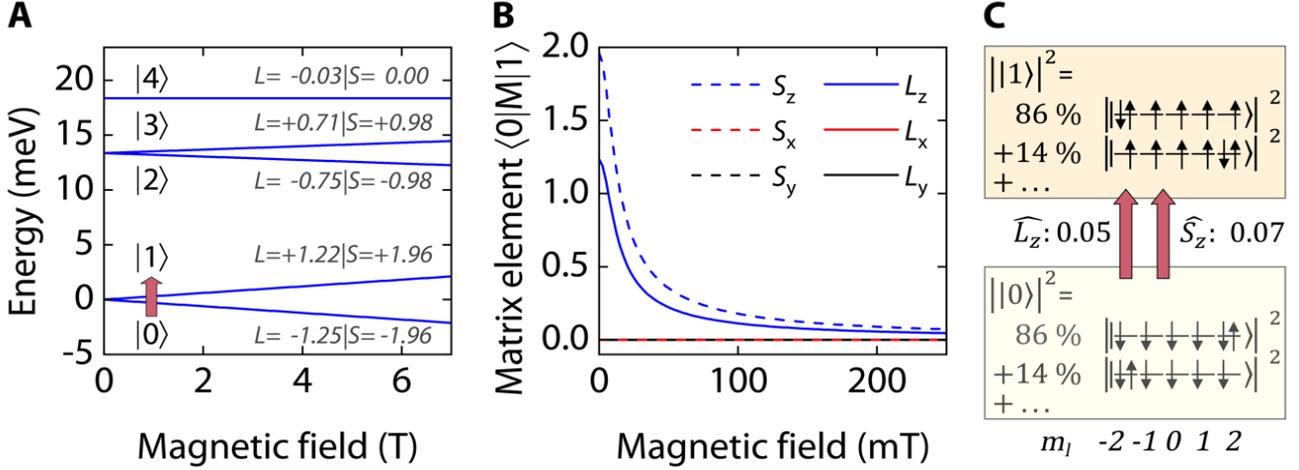

**Fig. 4. Energy levels and EPR materix elements of Fe/MgO/Ag(100). A** Calculated lowest energy levels of Fe obtained from the multiplet theory for an out-of-plane magnetic field ranging from 0 to 7 $T$. **B** Calculated components of the matrix elements of the orbital and spin momentum operator $\hat{L}$ and $\hat{S}$, respectively, for an external magnetic field along $z$. Note that apart from the operators $S_z$ and $L_z$, all other matrix elements are $< 10^{-14}$. **C** Schematic of the transition matrix elements between the EPR-active states $|0\rangle$ and $|1\rangle$ represented in the orbital-momentum basis $m_l$. Wave-function contributions below 1% are omitted.

The situation is more complex for Fe, which has a state multiplicity of 5 due to the effective spin $S = 2$, and the presence of strong orbital moments (see Fig. 4A). At zero magnetic field, the ground-state doublet is separated by about 14 meV from the next excited doublet (27). Therefore, only the two lowest states are thermally occupied in the range of temperature and magnetic field probed by our experiments. Transitions to higher doublets are too high in energy to be driven by the rf excitation. This renders Fe also an effective two-level system. Within this effective two-level system, we need to evaluate the interaction Hamiltonian $H'$ in the eigenstate basis $|0\rangle$ and $|1\rangle$, which are a general superposition of the $|M_L, M_S\rangle$ basis wave functions. To describe the states $|0\rangle$ and $|1\rangle$, we use the wave functions obtained from a multiplet model that was successfully employed to simultaneously describe the x-ray absorption spectral line shape and the low-energy excitations of Fe/MgO probed by STM (27). We find that the off-diagonal matrix elements $H'_{01}$ are proportional to $B_{1,z}$ whereas the in-plane field components, $B_{1,x}$ and $B_{1,y}$, yield vanishing matrix elements (see Materials and Methods and Section S6 for more details). In other words, only the $z$-component of the magnetic moment operator yields an off-diagonal matrix element $\langle 1|\hat{L}_z + 2\hat{S}_z|0\rangle \neq 0$.

This is known in EPR spectroscopy to be the case for integer spins where longitudinal $\mathbf{B}_1$ fields are used to drive the EPR transition (*29*). Further, the matrix element is strongly dependent on the Zeeman-splitting field $B_0$, as shown in Fig. 4B. Fe behaves as an integer-spin system, in which the levels are strongly intermixed by the crystal field as well as spin-orbit interaction (Fig. 4C). This leads to wave functions that are not eigenfunctions of the Zeeman Hamiltonian, thus the composition of the states $|0\rangle$ and $|1\rangle$ changes with the external magnetic field. Setting again $\hbar\Omega$ equal to the amplitude of $H'_{01}$, we see that the Rabi rate $\Omega$, besides being proportional to the $z$-component of the $\mathbf{B}_1$ field, is also proportional to the matrix element $\langle 1|\hat{L}_z + 2\hat{S}_z|0\rangle$.

Further, we describe the effective two-level system for Fe, not in terms of the magnetic-moment operator, but by the two-level polarization operator $\hat{\mathbf{P}} \propto \hat{\boldsymbol{\sigma}}$ (see Section S6). Following this approach, we can model the two EPR species using the same model for the tunneling current while taking into account that the origin of the Rabi rate is different for the two species. We derive the Bloch equations in terms of the polarization vector $\hat{\mathbf{P}}$ with a driving term proportional to $\hat{P}_x$ and with an effective driving field strength given by $\mu_B \langle 1|\hat{L}_z + 2\hat{S}_z|0\rangle B_{1,z}/\hbar$. Moreover, in the evaluation of the TMR for the read out of the EPR signal in the STM junction, we use the polarization vector $\hat{\mathbf{P}}$ instead of the physical magnetic moment of the system (see Section S2) since the conductance of the STM junction should only depend on the occupation and coherence of the involved EPR states (*26*). This approach reflects the fact that the conductance of the STM junction depends on the nature of the magnetic-adatom states and not only on the associated magnetic moment. Note that, for a real $S = 1/2$ system, the polarization operator is identical to the spin operator.

Thus, we obtain formally the same equation, Eq. 1, for the experimentally detected EPR signal for the two EPR species, Fe and TiH. However, the physical interpretation of the effective driving field component and strength that yield $\Omega$ differs for the two cases. In summary, our analysis shows that EPR in systems with $S > 1/2$ can be driven by STM, provided that longitudinal field gradients are nonzero. Note, that the small in-plane magnetic-field component of the tip produces negligible matrix elements for the in-plane magnetic moment operator as compared to its $z$-component (see Figure S8).

**Magnetic field acting on the adatoms**

At the position of the EPR species, the total magnetic field is the sum of the external magnetic field $\mathbf{B}_\text{ext}$ and the tip-induced effective magnetic field $\mathbf{B}_\text{eff}$. Quantitative analysis of the Rabi rate requires estimation of $\mathbf{B}_\text{eff}$ acting on the EPR species. Here, we determine $\mathbf{B}_\text{eff}$ by considering the measured resonance positions $B^0_\text{ext}$, i.e., the value of $\mathbf{B}_\text{ext}$ at resonance, as shown in Fig. 5 (A and B). The intrinsic resonance position in the absence of a tip-induced magnetic field is given by $2\hbar\omega_\text{rf}/\mu$, which yields 247 mT for Fe at $\omega^\text{Fe}_\text{rf}/2\pi = 36\,\text{GHz}$ and 286 mT for TiH at $\omega^\text{Ti}_\text{rf}/2\pi = 8\,\text{GHz}$. The measured EPR resonance position clearly deviates from these values as a function of the standoff distance $s$. These deviations are caused by the finite $\mathbf{B}_\text{eff}$ produced by the tip. Remarkably, the upturn of $B^0_\text{ext}$ at $s \approx 420\,\text{pm}$ for Fe indicates that the magnetic force changes from attractive to repulsive upon approaching this specific tip (Fig. 5A), which is unexpected for an interaction that only contains an exchange contribution as determined in previous studies (*2*, *7*, *8*). This finding indicates that the tip magnetic field comprises two competing terms, which we assume to be an exchange field $\mathbf{B}_{xc}$ and an additional dipolar field $\mathbf{B}_{dip}$. These two fields were shown to be present independently from one another for certain STM tips in Ref. (*6*) and were also discussed but not taken into account simultaneously in Ref. (*4*). We note that previous studies using an atomic-force microscope with a magnetic tip (*30*) have shown that the exchange interaction might change sign depending on the overlap of the tip and the magnetic adatom wave functions. Here, however, we find that the dipolar field in addition to an exponentially-decaying exchange field is sufficient to account for the observed change of $\mathbf{B}_{eff}$ without considering more complex exchange regimes. Given the cylindrical symmetry of the STM junction, it is sufficient to determine the $x$ and $z$ components of $\mathbf{B}_{eff}$, which can be written as (*22*)

$$\mathbf{B}_{eff} = \begin{pmatrix} B_{xc,x} + B_{dip,x} \\ B_{xc,z} + B_{dip,z} \end{pmatrix}$$
$$= \begin{pmatrix} \left(ae^{-s/\lambda_{xc}} - \dfrac{b}{s^3}\right)\sin\alpha \\ \left(ae^{-s/\lambda_{xc}} + \dfrac{2b}{s^3}\right)\cos\alpha \end{pmatrix}, \quad (2)$$

where $a$ is the exchange parameter, $b$ the dipolar parameter, $s$ the tip standoff distance defined through point-contact between the tip and the magnetic adatom (see Materials and Methods), and $\lambda_{xc}$ the exchange decay length. Note that we orient the coordinate system such that $B_{eff,y} = dB_{eff,y}/ds = 0$ along the $z$-axis and for $x = 0$. We fit $B_{ext}^0(s)$ using $B_{eff,z} + 2\hbar\omega_{rf}/\mu$ (see Eq. 2) with a fixed $\alpha = 64°$ as determined above, and find a good agreement between experiment and theory for $\lambda_{xc}^{Fe} = (370 \pm 60)$ pm, $a^{Fe} = (-0.6 \pm 0.1)$ T, $\lambda_{xc}^{Ti} = (170 \pm 20)$ pm, $a^{Ti} = (-2.2 \pm 0.1)$ T and $b = (0.2 \pm 0.03)\mu_0\mu_B$, see Fig. 5, A and B.

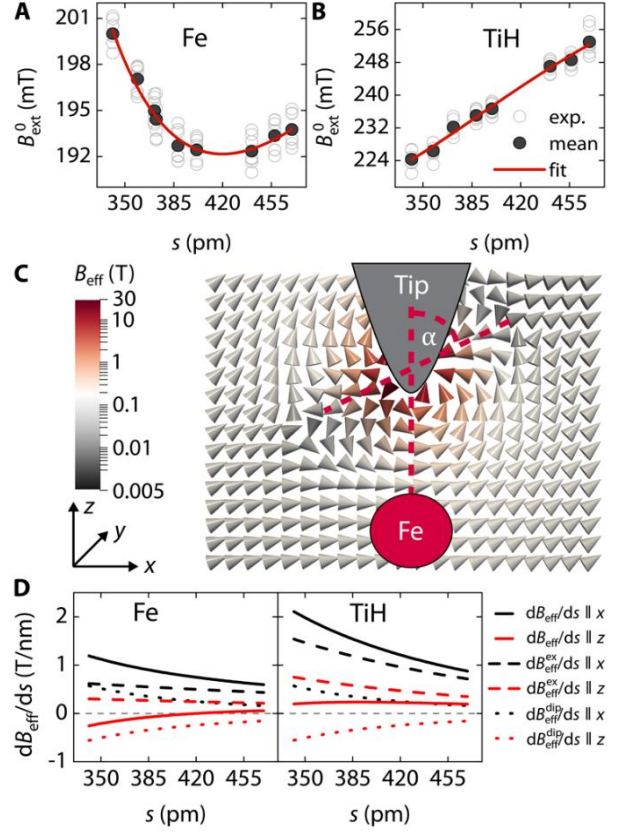

**Fig. 5. Characterization of the tip magnetic field.** Measured resonance field $B_{ext}^0$ vs standoff distance $s$ for Fe (**A**) and TiH (**B**). Solid lines are fits based on Eq. 2. (**C**) Cross sectional view of the effective tip magnetic field $\mathbf{B}_{eff}$ experienced by the Fe atom at different locations with respect to the STM tip deduced from Eq. 2 assuming an isotropic exchange interaction. The cross section is a cut along the tip-atom plane with dimensions 0.8x0.6 nm². (**D**) Gradient of the effective magnetic field $d\mathbf{B}_{eff}/ds$ along $x$ and $z$ vs standoff distance $s$ for Fe (left) and TiH (right) with the corresponding dipolar ($B_{dip}$) and exchange ($B_{xc}$) contributions. The gradients along $y$ vanish.

The parameter $b$ implies a tip magnetic moment of about 3 $\mu_B$, which is reasonable given that few Fe atoms form the tip apex. The values for $a$ compare well with reports of the exchange field, ranging from 0.1 to 10 T for similar systems (*4*, *7*, *31*). For all six tips used within this work (see Fig. S9), we find values of $a < 0$ independent of the adatom. The values for $\lambda_{xc}$ are somewhat larger than reported values (*4*, *7*, *31*) but $\lambda_{xc}$ is expected to strongly depend on the detailed atomic structure of the microtip. In this way, we completely characterize $\mathbf{B}_{\text{eff}}$ for this magnetic microtip, which is shown for Fe in Fig. 5C assuming an isotropic exchange interaction. This allows us to derive the corresponding magnetic-field gradients along $z$ as shown in Fig 5D containing significant contributions from dipolar and exchange tip-adatom interactions at the same time.

**Quantitative evaluation of the PEC Rabi rate and comparison with experiment**

Knowledge of the transition matrix elements and $\mathbf{B}_{\text{eff}}$ is essential to compute the Rabi rate expected for the PEC model [see Section S2 and Ref. (22)], which is given by

$$\Omega_{\text{PEC}}^{\text{Fe,TiH}} = \left| \frac{\mu_B}{\hbar} \Delta z_{\omega_{\text{rf}}} \frac{d\mathbf{B}_{\text{eff}z,x}}{ds} \langle 1|\hat{\mathbf{L}} + 2\hat{\mathbf{S}}|0\rangle \right|. \quad (3)$$

Here, different components of the magnetic-field gradient drive Fe and TiH as discussed above. In more detail, the field that drives EPR is given by $B_1^{\text{Fe}} = \Delta z_{\omega_{\text{rf}}}^{\text{Fe}} dB_{\text{eff}_z}^{\text{Fe}}/ds$ and $B_1^{\text{TiH}} = \Delta z_{\omega_{\text{rf}}}^{\text{TiH}} dB_{\text{eff}_x}^{\text{TiH}}/ds$, where $\Delta z_{\omega_{\text{rf}}}$ is the amplitude of the magnetic-adatom displacement induced by the rf electric field between tip and adatom. In order to compute $\Delta z_{\omega_{\text{rf}}}$, we calculate the structural response of the adatoms to a static electric field applied normal to the surface by means of density functional theory (DFT) (for details of the calculations see Materials and Methods). Since the adsorbate species Fe and TiH form a polar bond to the MgO substrate, an external electric field can displace the adatoms and vary the length of the bond to the surface. As the frequencies of the local vibrational modes of the adatoms lie at several THz [see Materials and Methods, Fig. S7 and previous work (7)], we expect $\Delta z_{\omega_{\text{rf}}}$ to adiabatically adjust to the GHz electric fields, justifying our static approach in the calculations. As seen in Fig. 6 (A and B), the Fe and Ti adatoms are both displaced by about 0.5 pm/(V/nm), but in opposite directions. The opposite response appears as a 180° phase change in the driving terms and has no consequences for the measurements. Note that the inverted EPR spectra of Fe and TiH originate from the sign change in $a_{\text{TMR}}$. Surprisingly, we observe that the displacement does not depend linearly on the electric field but follows a second-order polynomial (see also Fig. S7). This result is rationalized by noting that the linear response is due to the coupling to permanent electric dipoles, whereas the second-order term arises from a coupling to induced electric dipoles that has not been reported before (*7*). Finally, $\Delta z_{\omega_{rf}}$ is obtained by considering only the terms oscillating at the fundamental frequency $\omega_{rf}$ derived from the second-order polynomial fit of the displacement (see Fig. 6, A and B), i.e., neglecting time-independent offsets and terms oscillating at $2\omega_{rf}$, and using the experimental electric field $E = [V_{dc} + V_{rf}\cos(\omega_{rf}t)]/s$ (see Section S5).

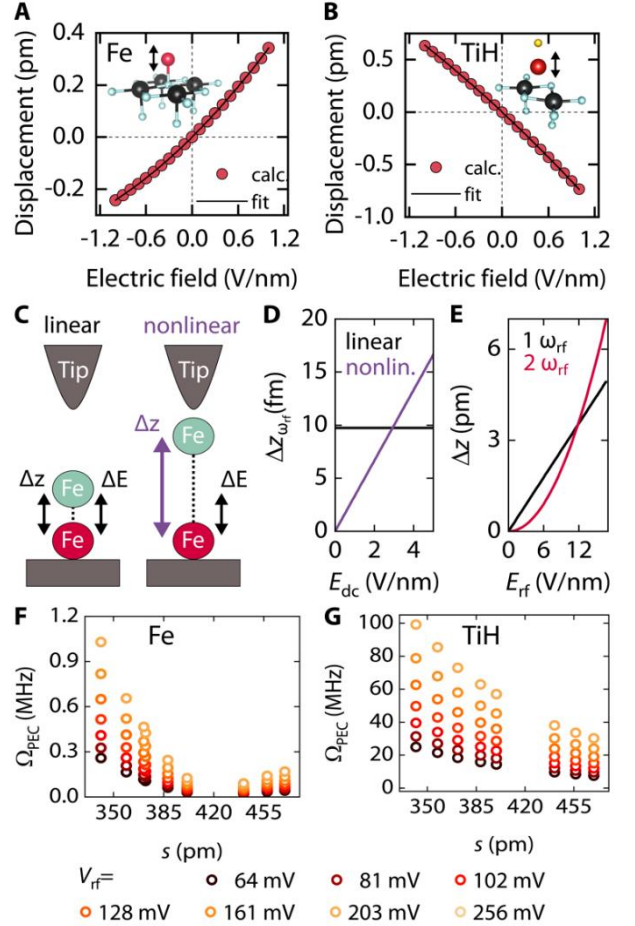

**Fig. 6. Adatom displacement induced by the electric field in the STM junction and calculated Rabi rate. A** (**B**) Detail of the atomic arrangement used for the DFT calculations of Fe (TiH) and calculated displacement vs static out-of-plane electric field. Color code: Mg (black), O (light-blue), Ti (red), H (yellow), Fe (purple). **C** Schematic of linear and nonlinear displacements $\Delta z$ due to the applied electric field $\Delta E$. **D** Calculated linear and nonlinear displacement at the rf frequency $\Delta z_{\omega_{\text{rf}}}$ vs dc electric field $E_{\text{dc}}$ for $V_{\text{rf}} = 10$ mV. **E** Calculated displacement $\Delta z$ vs rf electric field $E_{\text{rf}}$ for $V_{\text{dc}} = 10$ mV at the fundamental frequency $\omega_{\text{rf}}$ and the second harmonic frequency $2\omega_{\text{rf}}$ of the driving rf field. Standoff distance is 300 pm for D and E. **F** (**G**) Calculated Rabi rate $\Omega_{\text{PEC}}$ vs standoff distance $s$ for Fe (TiH) deduced from Eq. 3.

After these steps, we can finally compute the PEC Rabi rate $\Omega_{PEC}$ using Eq. 3, as shown in Fig. 6 (F and G). Remarkably, we find that the calculated values of $\Omega_{PEC}$ match the experimental Rabi rate $\Omega$ reported in Fig. 3 (E and F) for TiH and only deviate by a factor of 2 for Fe. Notably, $\Omega_{PEC}$ describes the $\Omega(s)$ dependence adequately for Fe, i.e., changing from decreasing to slightly increasing at $s \approx 420$ pm. This change in slope arises from the differences in the distance-dependence of the exchange and dipolar interactions with the adatom. Also for TiH, the decreasing trend of $\Omega$ with $s$ is reproduced correctly. Discrepancies in the magnitude are ascribed to an inaccurate determination of the electric field, which was shown to deviate from the plate-capacitor model used here (*32*). Moreover, keeping the adatom magnetic moment fixed along $z$ is especially critical for TiH at small standoff distances and can lead to errors. Lastly, including a finite phase between driving field and the precessing magnetic adatom spin, as well as a bias-dependent TMR and a spin-torque initialization (*5*) could further improve the agreement with the experiment.

## DISCUSSION

Given the limitations of our model, the overall good agreement between experiment and theory shows that the PEC mechanism allows for a consistent interpretation of EPR-STM spectra provided that the matrix elements of the EPR transitions and the different components of the magnetic field gradients are properly accounted for. Crucially, we find that in $S = 1/2$ systems, such as TiH, the rf magnetic field perpendicular to the magnetic moment drives EPR, whereas for the more complex $S = 2$ Fe system we find a different driving force, i.e., the rf magnetic field along the static magnetic field. This finding reflects the fact that transitions between states with spin quantum numbers $m_S = \pm 2$ as in Fe imply a change in spin-angular momentum of $4\hbar$ and are therefore dipole-forbidden, i.e., cannot be driven by a transverse rf magnetic field because the rf photons can only provide $1\hbar$. Instead, these transitions are enabled by the mixing of the ground and first excited state as found in Fe (*27*), which allows for a longitudinal rf magnetic field to drive EPR. Such distinct EPR driving forces for transitions that are magnetic-dipole forbidden have been also observed in ensemble EPR measurements (*29*), where they are known as longitudinal or parallel-polarization EPR.

The larger EPR amplitudes measured for TiH compared to Fe are mainly caused by the ten times larger EPR transition matrix element of TiH and the increasing weight of the homodyne-detection channel with increasing $V_{rf}$ as compared to Fe, where this detection channel is ineffective due to the larger $T_1$.

As mentioned previously, the PEC mechanism can be understood as a special case of an EPR theory involving a cotunneling picture (*23*). In that mechanism, the rf electric field alters the tunneling barrier that can be effectively mapped onto a time-dependent overlap of adatom and tip wavefunctions, which accounts also for a time-dependent exchange coupling. Thus, this model includes the PEC mechanism, in which the magnetic-adatom vibration causes the magnetic adatom-tip interactions to vary over time. Similarly, also the treatment of EPR-STM within an open-quantum-system approach (*26*) is not in contradiction with the PEC mechanism. This model accounts for the coupling of the EPR species and spin-polarized tip to reservoirs of energy and angular momentum and, additionally, introduces generalized Bloch equations to explain EPR, consistent with our treatment. However, this approach does not specifically address how the EPR transitions are driven, but rather outlines how they are sensed by the tunneling current in the experiment. Thus, these concepts can be combined with the theory used in this work to yield a full quantum description of EPR-STM in the future.

Our study also shows that EPR of single Fe atoms is possible at temperatures of 5 K using an out-of-plane external magnetic field, unlike in Refs. (*1, 5, 6, 8, 12*) that used predominantly in-plane fields (see also the discussion on the influence of the magnetic tip in Section S1 of the Supplementary Material). As indicated by our multiplet calculations, an in-plane magnetic field increases the EPR signal only very weakly (compare with Fig. 4, Section S6 and Fig. S8, A-C) and is not required in principle. In contrast, we find an optimal out-of-plane magnetic field of about 130 mT that is a compromise between the rapidly decreasing EPR transition matrix element for increasing an out-of-plane magnetic field and the off-resonant population difference between that states $|0\rangle$ and $|1\rangle$ that is proportional to $\tanh(\mu B_0/k_B T)$, where $k_B$ is the Boltzmann constant and $T = 4$ K. Note that spin pumping has been neglected and that considering additionally the dependence of the tip polarization on external field might further increase this optimal magnetic field.

In contrast to previous studies (*2, 7*), we show that the shift of the resonance magnetic field with the standoff distance is not determined by the orientation of the exchange field $\mathbf{B}_{xc}$ alone. In other words, the direction of the shift does not allow discriminating between antiferro- and ferromagnetic exchange coupling of the EPR species and tip. Instead, the shift direction is determined by the interplay between $\mathbf{B}_{xc}$ and the dipolar magnetic field as given in Eq. 2. Additionally, we find that the signs of the tunneling magnetoresistance and of the exchange field do not correlate, which might be caused by different contributing electronic states (*33*).

Our DFT modelling allows for a precise calculation of the magnetic-adatom displacements, which are about 0.1 pm at the rf frequency. More specifically, we find a displacement smaller by a factor of five for the Ti atom in the TiH system compared to previous calculations (*7*) at a standoff distance of 430 pm, a dc bias voltage of 50 mV and an rf-voltage amplitude of 10 mV. This difference highlights the importance of calculating the adatom displacement directly, i.e., without involving harmonic approximations of the computed energy landscape as a function of the external electric field. Importantly, we demonstrate in Fig. 6 (C and D), how the Rabi rate can be tuned by the dc bias voltage through coupling to induced changes in electric dipoles, which readily account for up to 15% of the Fe displacement in our experiment (see also Section S5). In Fig. S7 we report additional DFT calculations of the displacement of Fe at larger electric fields that show its strong nonlinear response and highlight again the profound impact of induced electric dipole moments on the magnetic-adatom displacement. Such a nonlinear response should also enable the driving of EPR at the second harmonic frequency of the rf field (see Fig. 6E). For experimental parameters that are within reach in future studies, both of these predicted nonlinear driving mechanisms (second-harmonic driving at $V_{dc} = 10$ mV, $V_{rf} = 3$ V and $s = 300$ pm, induced electric dipoles at $V_{dc} = 1$ V, $V_{rf} = 10$ mV and $s = 300$ pm) outperform their linear counterparts as shown in Fig 6 (D and E) underlining their potential to drive EPR-STM in a broader range of systems than demonstrated to date.

In summary, our combined experimental and theoretical investigation provides a consistent picture of EPR-STM of transition-metal adatoms on MgO. Our analysis also allows for fully characterizing the vector magnetic field of the tip, which is convenient for future EPR-STM studies and other STM studies relying on spin-polarized tunneling (*34*). Whereas EPR-STM measurements have been so far only reported for transition-metal atoms on MgO, the observation of adatom displacements under rf excitation arising from induced electric dipoles opens the field of EPR-STM to nonpolar paramagnetic species. Moreover, our conclusions suggest that non-resonant EPR driving via second-harmonic generation might be feasible, thus allowing for strict separation of the excitation from the probe in pulsed EPR studies (*12*). Such nonlinear driving could also enhance the coupling efficiency when approaching the resonant THz frequency of phonons by an rf-photon up-conversion scheme, which will additionally benefit from reduced losses in signal transmission at lower rf frequencies.

## MATERIALS AND METHODS

### Experimental Design

Measurements are performed using a Joule-Thomson STM from Specs operating at 4.5 K and upgraded for rf capabilities [see Fig. 1A and Ref. (*10*)]. $V_{rf}$ is characterized by rectification at a STM-junction-conductance nonlinearity (see below). The sample is a clean Ag(100) surface on which double-layer MgO islands are grown (Fig. 1B) (*10*). Single Fe and Ti atoms are deposited on the cold sample inside the STM. Residual $H_2$ gas is known to hydrogenate Ti forming TiH complexes (*2, 10*). The tip is made from a chemically etched W wire that is dipped into the sample to obtain a sharp apex. Spin contrast is achieved by picking up single Fe atoms. We check for tip changes by scanning the respective area before and after EPR spectra were recorded and by recording an EPR spectrum at the beginning and at the end of a parameter sweep with the same settings. The standoff distance $s$ is calibrated by point-contact measurements, $I_{dc}(z)$ and $z(V_{dc})$ curves (see below).

EPR spectra are acquired by sweeping the out-of-plane $\mathbf{B}_{ext}$ at a constant rf frequency with a sweep rate of 400 µT/s. We modulate the rf voltage with a square wave at 971 Hz and recording the first harmonic of the tunneling current $\Delta I$ at the modulation frequency using a lock-in amplifier (LIA). During EPR sweeps, the tip circulates above the EPR species at a rate of 383 Hz with a radius of 10 pm to track the adatom. The systematic spread in $B_0$ for constant $s$ of about $\pm 1$ mT [see Fig. 3, C and D] arises from opposite $B_{ext}$-sweep directions and the limited Hall-probe communication speed.

We choose EPR species separated from other magnetic adatoms by more than 3 nm to minimize magnetic interactions (see Fig. 1B). All EPR sweeps on TiH are recorded on the bridge binding site with respect to the oxygen sublattice; notably, TiH on the oxygen binding site quickly destabilizes upon rf excitation. For each EPR sweep, a nonresonant reference spectrum is recorded and substracted (see below).

$dI/dV$ spectroscopy is performed by adding a sinusoidal voltage (971 Hz, amplitude of a few millivolts) to the dc bias and using a lock-in technique.

### Characterization of $V_{rf}$

We characterize the rf-voltage amplitude at the STM junction by rectification of the rf signal at a STM-junction-conductance nonlinearity as outlined in Ref. (*10*). This procedure is performed at the two frequencies used for EPR sweeps, i.e., at 8 GHz and at 36 GHz (see Fig. S1, A and B).

### Characterization of the standoff distance $s$

The standoff distance is characterized in three steps:

(i) We perform point-contact measurements in which we open the feedback at 10 mV dc bias and approach with the tip while recording the dc current. At point contact, a plateau in dc current is reached (see Fig. S2, A and B). The extracted point-contact conductances are consistent with reported values for Fe (*28*) and a bridge-binding-site TiH adatom (*4*). From this measurement, we calibrate the absolute tip height above the adatom. Since the value of conductance at point contact was found to be independent of the microtip to a good approximation, we do not repeat this measurement for each microtip used for EPR because it has a high risk of altering the microtip. Such similar conductivity at point contact for different microtips can be expected given the fact that the adatom-MgO-Ag junction is the current-limiting part.

(ii) We record $I(z)$ curves for the specific microtip used for EPR sweeps avoiding point contact with a finer resolution than in (i) in the range of interest for the EPR spectra. With the point- contact measurement of (i) and by fitting the data with an exponential, the absolute standoff distance is determined (see Fig. S2, C and D).

(iii) We perform $z(V)$ measurements for the specific microtip used for EPR sweeps opening the feedback at the values of $I_{dc}$ used in the EPR sweeps to account for the rigid shift in standoff distance upon change of $V_{dc}$ (see Fig. S2, E and F).

We note that steps (ii) and (iii) are performed at about 200 mT external magnetic field to match the EPR experimental conditions.

**Characterization of the rf current**

To account for the rf-current-induced relaxation processes correctly, the rf-current amplitude has to be characterized. Ideally this is done via convoluting the experimental $dI/dV$ curve with a sinusoidal rf voltage of the corresponding amplitude over one period. However, this requires a detailed knowledge of the $dI/dV$ curve, which changes with the setpoint and the external magnetic field. In Figure S3A, we compare this approach to an approximation, in which the rf-current amplitude is computed via Ohm's law using the dc tunneling resistance at the setpoint. From the very good agreement between the two approaches, we conclude that the latter approach is also valid. Note that the data in Fig. S3A is obtained for an additional EPR data set on TiH for varying $I_{dc}$, $V_{dc}$ and $V_{rf}$ shown in Fig. S9A.

**EPR reference spectra**

The background signals in EPR sweeps are caused by rectification of the rf voltage at STM-junction-conductance nonlinearities (*10*). Some of these nonlinearities are of magnetic origin. This means that they change if either the tip or the adatom change their magnetic polarization. Since our EPR sweeps are performed in field ranges, where neither the adatom nor presumably the tip are fully spin polarized, the rf rectification will depend on the external field. On the other hand, the STM-junction conductance also strongly depends on $I_{dc}$, $V_{dc}$ and clearly on $V_{rf}$. In order to account for changes in the conductance nonlinearities, i.e. a change of the tip and atoms magnetic polarization, as we sweep the magnetic field, a nonresonant background signal is recorded for each of the 119 EPR spectra. For the Fe adatom, a reference sweep at a constant rf frequency of 8 GHz is performed (see Fig. S3B) that we subtract from the resonant sweep at 36 GHz. To this end, the rf-voltage amplitude at 8 GHz is matched to the one at 36 GHz by compensating for the rf transfer function towards the STM junction. For TiH, a similar procedure is applied, but the reference is recorded at 36 GHz, whereas the resonant sweep is performed at 8 GHz (see Fig. S3C). Note that for the largest values of $V_{rf}$, a minor inaccuracy in compensating for the rf-transfer function required a rescaling of the reference spectrum by a constant that is close to unity to best match the background of the resonant EPR spectrum before subtraction.

## Details of the fit procedure

The best fit of the 119 EPR spectra (see Fig. 2 and Fig. S4) to Eq. 1 is obtained by minimizing the root mean square deviation from the normalized EPR signal given by $(\Delta I - I_{\text{off}})/I_{\text{dc}}$. This accounts for the anticipated large dynamic range in $\Delta I$ as a function of $I_{\text{dc}}$, i.e., to improve the fit accuracy for small $I_{\text{dc}}$, for which our model assumptions are most appropriate (see discussion in the main text concerning the moving adatom spin angle at closest distances, i.e. for large $I_{\text{dc}}$).

Further, our model employs in total 126 parameters that determine the spectral line shape for 119 EPR spectra, i.e., on average 1.06 free parameter per spectrum. This demonstrates that we chose a minimized set of parameters considering that a Lorentzian line shape is in principle determined by 3 parameters. See also the discussion on the number of fit parameters in Section S2 of the Supplementary Materials.

To determine the global minimum of the fit, we vary the starting conditions and take the result with the smallest root-mean-square deviation. Figure S5 shows the resulting deviations for different starting parameters of α. We note that our model yields $T_1$ times that are larger than reported in previous studies (*5, 28*), in which an in-plane component of the external field of about 10% was present. Additionally, the fact that we assume an atom-tracking-induced additional broadening independent of the EPR species can lead to an apparent increase in $T_1$ in the fit as we verified by additional tests. Our model also neglects relaxation mediated by phonons due to the relatively high tunneling currents and the thin MgO support (*28*).

We determine the uncertainties in the fit parameters related to Eq. 2 by standard error analysis. For the fit parameters related to Eq. 1, this approach is hampered by the complexity of the fit procedure. Therefore, we first determine the average experimental noise to signal ratio to be 2%. In the next step, we vary each fit parameter related to Eq. 1 separately until the root-mean-square deviation of the fit from the experimental spectra grows by 2%. For the Rabi rates, we vary all 119 values at once by an absolute value until the latter 2% deviation is observed.

## Additional data sets

Importantly, our conclusions are consistent with several additional data sets acquired for a similar range of parameters that we show in Fig. S9.

## Multiplet calculations

The Fe wave functions, corresponding properties and matrix elements are obtained from charge-transfer multiplet calculations. The crystal-field and charge-transfer parameters are taken from previous calculations for the simulation of x-ray absorption spectra of the same system (*27*). In this model, the Fe adatom is described by a combination of $d^6$ and $d^7$ configurations coupled by a charge-transfer term, in which an electron from a filled substrate oxygen-derived shell is allowed to hop onto the d-shell of Fe via the $d_{z^2}$ orbital. The Slater-Condon integrals are rescaled to 75% of their Hartree-Fock value and the one-electron spin-orbit coupling constant of Fe is taken to be 52 meV for the $d^6$ and 45 meV for the $d^7$ configuration. The charge-transfer energy between the $d^6$ and $d^7$ configurations amounts to 0.5 eV where the hopping parameter to the $d_{z^2}$ orbital is 0.85 eV. The crystal field is chosen to be the same for the $d^6$ and for the $d^7$ configuration and is given by $10D_q = -0.13$ eV, $D_s = -0.44$ eV and $D_t = -0.015$ eV.

**Density functional theory calculations**

For our first-principles calculations we use the density functional theory formalism as implemented in the Vienna ab-initio simulation package (VASP) (*35*). For the Fe adatom, we use a 49-atom unit cell with Fe located above a surface oxygen. For the TiH adatoms, we use a 50-atom unit cell with TiH located above a surface oxygen-oxygen bridge. We top both unit cells by 16 Å of vacuum to achieve convergence of forces, and we fix the in-plane lattice constant of the bottom MgO layer to that of Ag(100) (289 pm). Since MgO was shown to act as an efficient filter for the phonon modes of a substrate (*36*), we do not take the Ag substrate into account in this calculation. We use the default VASP PAW pseudopotentials and converge the Hellmann-Feynman forces to $10^{-5}$ eV/Å using a plane-wave energy cut-off of 750 eV and a $3 \times 3 \times 1$ k-point mesh to sample the Brillouin zone. For the exchange-correlation functional we choose the PBEsol form of the generalized gradient approximation (*37*). Our fully relaxed structure with a MgO in-plane lattice constant of 291 pm fits reasonably well to the experimental values of Ref. (*36*). The Fe adatom is elevated 194 pm above the protruded surface oxygen. The Ti in the TiH system adatom is elevated 198 pm above the surface oxygen-oxygen bridge, and the bond length of the TiH molecule is 177 pm. An illustration of the unit cells is shown in Fig. S7 (A and B).

We calculate the vibrational frequencies and eigenvectors using the frozen-phonon method as implemented in the phonopy package (*38*). The calculations reveal low-frequency localized vibrational modes at the Brillouin zone center involving mainly the motion of the Fe adatom parallel to the surface around 1.9 THz and perpendicular to the surface around 2.9 THz. We obtain the main contributions of the TiH molecule to the vibrational spectrum between 2 and 4 THz and one intramolecular vibrational mode around 10 THz. These modes show up in the vibrational density of states as peaks in the low-frequency regime, as shown in Fig. S7 (C and D). Vibrational modes involving mainly the ions of the MgO slab lie at higher frequencies above roughly 5 THz.

We further calculate the Born effective charges using density functional perturbation theory (*39*). In absence of any external electric field, the diagonal component normal to the MgO surface is +0.32 *e* for Fe, +0.61 *e* for Ti, and -0.43 *e* for H, where *e* is the elementary charge.

Next, we model the structural changes of the systems in an applied electric field. The rf electric field used in the experiment is so in low frequency that we expect no excitation of phonons to occur. Instead, we expect the atoms to follow the electric field adiabatically. We therefore apply electric fields with different magnitudes between -1 V/nm and 1 V/nm normal to the MgO surface and relax the atomic positions to estimate the induced relative shifts of the Fe and TiH adatoms. The results are shown in Fig. 6 (A and B) and in Fig. S7 (E and F).

**Acknowledgments**

**Funding:** This work was supported by the Swiss National Science Foundation, project # 200021_163225. DMJ received support from the Swiss National Science Foundation (SNSF) under Project ID 184259. Density functional theory calculations were performed at the Swiss National Supercomputing Centre (CSCS) supported by the Project ID s624 and p504.

**Author contributions:** TSS, SK, PG and SS conceived the experiment. TSS and SK performed the measurements. DMJ and NAS performed the density functional theory calculations. SS developed the density-matrix formalism for the Bloch equations and performed the multiplet calculations. TSS analyzed the data and wrote the manuscript with assistance from SK, PG and SS. All authors commented on the manuscript.

**Competing interests:** The authors declare no competing interests.

**Data and materials availability:** Data and materials included in this work are available on reasonable request.

# SUPPLEMENTARY MATERIALS

Section S1. Role of the magnetic tip in EPR-STM

Section S2. Derivation of the EPR-STM signal

Section S3. Detailed EPR data set

Section S4. Additional fit results

Section S5. Fit of the calculated displacement and linearization in the driving electric field

Section S6. Details on charge-transfer multiplet calculations

Fig. S1. Characterization of the rf-voltage amplitude

Fig. S2. Characterization of the standoff distance

Fig. S3. Characterization of additional EPR parameters

Fig. S4. Detailed EPR data set with fits

Fig. S5. Details on fit procedure

Fig. S6. Additional EPR-spectra fit results

Fig. S7. Details on DFT calculations

Fig. S8. Details on multiplet calculations

Fig. S9. Additional EPR-STM data sets

Table S1. Ranges of fit parameters

## SECTION S1. ROLE OF THE MAGNETIC TIP IN EPR-STM

An important question in EPR-STM is why not every spin-polarized tip with proven magnetic contrast does not automatically yield EPR lines. To sense EPR on both Fe and TiH, α should differ from 0° and 90°. This precondition stems from the largely different $T_1$ of the two species, which requires the ability to detect both the dc and homodyne EPR signals (see Eq. 1). Accordingly, the data in Fig. 2 have been recorded with a microtip having $α = (64 \pm 2)°$. Although the exact microstructure of the STM tip is unknown, the deviation of the tip magnetic moment from $α = 0°$, which would be favored by tips having weak or axial magnetic anisotropy, indicates that local magnetic anisotropy at the apex counteracts $\mathbf{B}_{ext}$. As the tip apex structure can hardly be controlled or even characterized in STM, it is not surprising that the EPR sensitivity varies greatly from one tip to another. Indeed, out of more than 100 microtips tested in our setup, about 10% of the microtips yield EPR on either of the two adatoms but only about 1% work for both. This observation shows that, in addition to α, other factors affect the sensitivity of our measurements. In particular, the tunnel magnetoresistance of the junction $a_{TMR}$ has to be large enough for both adatoms. For the PEC mechanism, the effective tip magnetic field has to be sufficiently inhomogeneous along different directions for Fe and TiH, as shown further below. Lastly, fluctuating magnetic moments at the tip apex (5) must not decrease $T_2$ to a level that precludes sensing of the EPR signal.

## SECTION S2. DERIVATION OF THE EPR-STM SIGNAL

### Generalization of the Bloch equations in terms of density matrices

We start from the density matrix of a simple two-level system,

$$\hat{\rho} = \begin{pmatrix} \rho_{++} & \rho_{+-} \\ \rho_{-+} & \rho_{--} \end{pmatrix}. \quad (S1)$$

The density matrix is written in the basis states $\{|+\rangle, |-\rangle\}$, which are the eigenstates of the complete Hamiltonian including the state-splitting magnetic field $B_0$ that is given in the experiment by the sum of $B_{ext_z}$ and $B_{eff_z}$. The time evolution of the density matrix is given by

$$\frac{d}{dt}\hat{\rho} = -\frac{i}{\hbar}[\hat{H}, \hat{\rho}], \quad (S2)$$

where $\hbar$ is the reduced Planck constant. This describes only the coherent part of the time evolution, which is unitary, and does not include decoherence or relaxation due to interaction with the environment. The interaction of a quantum state that is generally described by a superposition of states (here $\{|+\rangle, |-\rangle\}$) destroys the state, i.e. it loses its coherence. Regarding the density matrix the interaction will reduce its coherence terms, i.e. the off-diagonal terms, which can be described by

$$\frac{d}{dt}\hat{\rho} = -\frac{1}{T_2}\begin{pmatrix} 0 & \rho_{+-} \\ \rho_{-+} & 0 \end{pmatrix}, \quad (S3)$$

where $T_2$ is called the phase coherence time or transversal relaxation time.

The environment can be described as a bath being in equilibrium at a certain temperature. The backaction of the quantum system on the bath is assumed to be negligible. The interaction of the bath with our quantum system tends to thermalize the quantum system towards the thermal state $\hat{\rho}_{\text{thermal}}$. This part of the relaxation is described by

$$\frac{d}{dt}\hat{\rho} = -\frac{1}{T_1}\left(\begin{pmatrix} \rho_{++} & 0 \\ 0 & \rho_{--} \end{pmatrix} - \hat{\rho}_{\text{thermal}}\right), \quad (S4)$$

Where $T_1$ is called the energy or longitudinal relaxation time. The thermal state is given as $\hat{\rho}_{\text{thermal}} = p_+^{\text{th}}|+\rangle\langle+| + p_-^{\text{th}}|-\rangle\langle-|$ and the $p_i^{\text{th}}$ are given by the Boltzmann distribution. Putting all the terms together, we obtain the Bloch equations

$$\frac{d}{dt}\hat{\rho} = -\frac{i}{\hbar}[\hat{H},\hat{\rho}] - \frac{1}{T_1}\left(\begin{pmatrix} \rho_{++} & 0 \\ 0 & \rho_{--} \end{pmatrix} - \hat{\rho}_{\text{thermal}}\right) - \frac{1}{T_2}\begin{pmatrix} 0 & \rho_{+-} \\ \rho_{-+} & 0 \end{pmatrix}. \quad (S5)$$

For a two-level system we can write the density matrix as

$$\hat{\rho} = \frac{I + \vec{n}\cdot\vec{\sigma}}{2}, \quad (S6)$$

where $\vec{n}$ describes the direction and amplitude of the polarization in the chosen coordinate system. Polarization has a generalized meaning and assumes that the states $\{|+\rangle, |-\rangle\}$ have a different polarization of some kind. The components are

$$\vec{n} = (2\text{Re}(\rho_{+-}), 2\text{Im}(\rho_{+-}), \rho_{++} - \rho_{--}). \quad (S7)$$

$\vec{\sigma} = (\sigma_x, \sigma_y, \sigma_z)$ are the standard Pauli matrices and $I$ is the $(2\times 2)$ identity matrix.

The expectation value of any operator can be obtained using

$$\langle \hat{S} \rangle = \text{Tr}[\hat{S}\hat{\rho}]. \quad (S8)$$

This entails also the time evolution of the operators,

$$\frac{d}{dt}\langle \hat{S} \rangle = \text{Tr}\left[\left(\frac{d}{dt}\hat{S}\right)\hat{\rho}\right] = \text{Tr}\left[\hat{S}\left(\frac{d}{dt}\hat{\rho}\right)\right]$$

$$= \text{Tr}\left[\hat{S}\left(-\frac{i}{\hbar}[\hat{H},\hat{\rho}] - \frac{1}{T_1}\left(\begin{pmatrix} \rho_{++} & 0 \\ 0 & \rho_{--} \end{pmatrix} - \hat{\rho}_{\text{thermal}}\right) - \frac{1}{T_2}\begin{pmatrix} 0 & \rho_{+-} \\ \rho_{-+} & 0 \end{pmatrix}\right)\right]$$

$$= -\frac{i}{\hbar}\text{Tr}\left[\hat{S}[\hat{H},\hat{\rho}]\right] - \frac{1}{T_1}\left(\text{Tr}\left[\hat{S}\begin{pmatrix} \rho_{++} & 0 \\ 0 & \rho_{--} \end{pmatrix}\right] - \text{Tr}[\hat{S}\hat{\rho}_{\text{thermal}}]\right) - \frac{1}{T_2}\text{Tr}\left[\hat{S}\begin{pmatrix} 0 & \rho_{+-} \\ \rho_{-+} & 0 \end{pmatrix}\right]$$

$$= -\frac{i}{\hbar}\langle[\hat{H},\hat{S}]\rangle - \frac{1}{T_1}\left(\langle\hat{S}\rangle_{\text{longitudinal}} - \langle\hat{S}\rangle_{\text{thermal}}\right) - \frac{1}{T_2}\langle\hat{S}\rangle_{\text{decoherence}}, \quad (S9)$$

where we used in the last step that the trace of a matrix is invariant under cyclic permutation.

The Hamiltonian of the system is given by the full system including the $B_0$ field, i.e. $\hat{H}_0$, and the small perturbation caused by the time-dependent $B_1(t)$ field, $\hat{H}'$. $\hat{H}_0$ is diagonal in the basis $\{|+\rangle, |-\rangle\}$,

$$\hat{H}_0 = \begin{pmatrix} E_+ & 0 \\ 0 & E_- \end{pmatrix}. \quad (S10)$$

The perturbation is given by the action of a small oscillating magnetic field $\mathbf{B}_1(t)$ acting on the magnetic moment of the system. We assume that the perturbating field does not change the character of the states $\{|+\rangle, |-\rangle\}$ themselves.

**Bloch equations for the TiH-MgO system**

Now, we want to consider a simple two-level spin system with the basis states $\{|\uparrow\rangle, |\downarrow\rangle\}$. The different parts of the Hamiltonians read

$$\hat{H}_0 = \begin{pmatrix} E_\uparrow & 0 \\ 0 & E_\downarrow \end{pmatrix} = \frac{1}{2}\hbar\omega_\mathrm{L} \begin{pmatrix} +1 & 0 \\ 0 & -1 \end{pmatrix} = \omega_\mathrm{L}\hat{S}_z, \qquad \hat{H}' = -\gamma\hat{\mathbf{S}}\cdot\mathbf{B}_1. \quad (S11)$$

The Lamor frequency $\omega_\mathrm{L}$ is given by $\omega_\mathrm{L} = \frac{g\mu_\mathrm{B}}{2\hbar}B_0$ with the g-factor $g$, i.e., $g = 2$ for a spin ½ system, and the Bohr magneton $\mu_\mathrm{B}$.

The gyromagnetic ratio for a spin ½ system is given by $\gamma = -\frac{g\mu_\mathrm{B}}{\hbar} = -2\frac{\mu_\mathrm{B}}{\hbar}$. In this basis, the components of $\hat{\mathbf{S}} = \frac{\hbar}{2}(\sigma_x, \sigma_y, \sigma_z)$ can be expressed in the basis of the Pauli spin matrices $\boldsymbol{\sigma}$,

$$\hat{S}_x = \frac{\hbar}{2}\begin{pmatrix} 0 & 1 \\ 1 & 0 \end{pmatrix}, \quad \hat{S}_y = \frac{\hbar}{2}\begin{pmatrix} 0 & -i \\ i & 0 \end{pmatrix}, \quad \hat{S}_z = \frac{\hbar}{2}\begin{pmatrix} 1 & 0 \\ 0 & -1 \end{pmatrix}. \quad (S12)$$

This means that this form of the spin operators is only given in the $\{|\uparrow\rangle, |\downarrow\rangle\}$ basis. For any other arbitrary basis $\{|+\rangle, |-\rangle\}$ the components of the $2 \times 2$ matrices need to be evaluated. For the time evolution of the spin operator without the $\mathbf{B}_1(t)$ field, we find

$$\frac{d}{dt}\langle\hat{\mathbf{S}}\rangle = -\frac{i}{\hbar}\langle[\hat{H}_0, \hat{\mathbf{S}}]\rangle - \frac{1}{T_1}\left(\langle\hat{\mathbf{S}}\rangle_\mathrm{longitudinal} - \langle\hat{\mathbf{S}}\rangle_\mathrm{thermal}\right) - \frac{1}{T_2}\langle\hat{\mathbf{S}}\rangle_\mathrm{decoherence}. \quad (S13)$$

For $= -\frac{i}{\hbar}\langle[\hat{H}_0, \hat{\mathbf{S}}]\rangle$ we can write now $= -i\frac{\omega_\mathrm{L}}{\hbar}\langle[\hat{S}_z, \hat{\mathbf{S}}]\rangle$. The components read then

$$\frac{d}{dt}\langle\hat{S}_x\rangle = \omega_\mathrm{L}\langle\hat{S}_y\rangle - \frac{1}{T_2}\langle\hat{S}_x\rangle \quad (S14)$$

$$\frac{d}{dt}\langle\hat{S}_y\rangle = -\omega_\mathrm{L}\langle\hat{S}_x\rangle - \frac{1}{T_2}\langle\hat{S}_y\rangle \quad (S15)$$

$$\frac{d}{dt}\langle\hat{S}_z\rangle = -\frac{1}{T_1}\left(\langle\hat{S}_z\rangle - \langle\hat{S}_z\rangle_\mathrm{thermal}\right). \quad (S16)$$

The expectation value is always understood as taking the trace of the operator with the density matrix of the system. Now, we turn on $\hat{H}'$ and need to evaluate in addition the term

$$-\frac{i}{\hbar}\langle[\hat{H}', \hat{\mathbf{S}}]\rangle = i\frac{\gamma}{\hbar}\langle[\hat{\mathbf{S}}\cdot\mathbf{B}_1, \hat{\mathbf{S}}]\rangle = -\gamma\langle\hat{\mathbf{S}}\rangle \times \mathbf{B}_1(t). \quad (S17)$$

Usually, $\mathbf{B}_1$ is said to be aligned perpendicular to $\mathbf{B}_0$ and to be linearly polarized in the $x$−direction, $\mathbf{B}_1(t) = \vec{e}_x B_1 \cos \omega t$, where $B_{1x} \equiv B_1$. To solve the above equations more easily, the linear polarized field is split into two right and left circular oscillating fields,

$$\mathbf{B}_1(t) = \left(\vec{e}_x \frac{B_1}{2} \cos \omega t + \vec{e}_y \frac{B_1}{2} \sin \omega t\right) + \left(\vec{e}_x \frac{B_1}{2} \cos \omega t - \vec{e}_y \frac{B_1}{2} \sin \omega t\right). \quad (S18)$$

The left rotating part (second term above) will be omitted, since it rotates counterclockwise to the Larmor-precession. Thus, we obtain for

$$-\gamma \langle \hat{\mathbf{S}} \rangle \times \mathbf{B}_1(t) = -\gamma \frac{B_1}{2} \begin{pmatrix} \langle \hat{S}_x \rangle \\ \langle \hat{S}_y \rangle \\ \langle \hat{S}_z \rangle \end{pmatrix} \times \begin{pmatrix} \cos \omega t \\ \sin \omega t \\ 0 \end{pmatrix} = \frac{\omega_1}{2} \begin{pmatrix} -\langle \hat{S}_z \rangle \sin \omega t \\ \langle \hat{S}_z \rangle \cos \omega t \\ \langle \hat{S}_x \rangle \sin \omega t - \langle \hat{S}_y \rangle \cos \omega t \end{pmatrix}, \quad (S19)$$

where we set $\omega_1 = -\gamma B_1$ ($\gamma < 0$ such that $\omega_1 > 0$). Using this result in the equation of motion above we eventually obtain

$$\frac{d}{dt} \langle \hat{S}_x \rangle = \omega_L \langle \hat{S}_y \rangle - \frac{\omega_1}{2} \sin \omega t \langle \hat{S}_z \rangle - \frac{1}{T_2} \langle \hat{S}_x \rangle \quad (S20)$$

$$\frac{d}{dt} \langle \hat{S}_y \rangle = -\omega_L \langle \hat{S}_x \rangle + \frac{\omega_1}{2} \cos \omega t \langle \hat{S}_z \rangle - \frac{1}{T_2} \langle \hat{S}_y \rangle \quad (S21)$$

$$\frac{d}{dt} \langle \hat{S}_z \rangle = \frac{\omega_1}{2} \left(\sin \omega t \langle \hat{S}_x \rangle - \cos \omega t \langle \hat{S}_y \rangle\right) - \frac{1}{T_1} \left(\langle \hat{S}_z \rangle - \langle \hat{S}_z \rangle_{\text{thermal}}\right). \quad (S22)$$

These are the standard equation one finds for the analysis of the behavior of a spin in an external magnetic field with a crossed oscillating magnetic field. Note the factor $1/2$ due to the decomposition of the linear polarized magnetic field $\mathbf{B}_1$ into the two counterclockwise rotating fields. This means that we identify the Rabi rate $\Omega = \frac{\omega_1}{2} = \frac{1}{2} \frac{g \mu_B}{\hbar} B_{1x} = \frac{\mu_B}{\hbar} B_{1x}$.

**Bloch equations for the Fe-MgO system**

Now, we want to modify two things in the derivation above. First, we will work now in the basis of the eigenstates of the Fe-MgO system, for which the expectation values for the magnetic moment operator can be evaluated, i.e. $\langle \pm | \hat{\mathbf{L}} + 2\hat{\mathbf{S}} | \pm \rangle$ and $\langle \pm | \hat{\mathbf{L}} + 2\hat{\mathbf{S}} | \mp \rangle$. In this basis, the unperturbed Hamiltonian $\hat{H}_0$ is still diagonal, i.e.

$$\hat{H}_0 = \begin{pmatrix} E_+ & 0 \\ 0 & E_- \end{pmatrix}. \quad (S23)$$

To evaluate the equation of motion we need to compute the matrices for $\hat{\mathbf{L}} + 2\hat{\mathbf{S}}$ in the basis of $\{|+\rangle, |-\rangle\}$. From the multiplet calculations [see Section S6 and (27)], we find that

$$\langle +|\hat{L}_x + 2\hat{S}_x|+\rangle = \langle -|\hat{L}_x + 2\hat{S}_x|-\rangle = \langle +|\hat{L}_x + 2\hat{S}_x|-\rangle = 0 \quad (S24)$$

$$\langle +|\hat{L}_y + 2\hat{S}_y|+\rangle = \langle -|\hat{L}_y + 2\hat{S}_y|-\rangle = \langle +|\hat{L}_y + 2\hat{S}_y|-\rangle = 0 \quad (S25)$$

$$-\langle +|\hat{L}_z + 2\hat{S}_z|+\rangle = \langle -|\hat{L}_z + 2\hat{S}_z|-\rangle = -\hbar m, \quad \langle +|\hat{L}_z + 2\hat{S}_z|-\rangle = \hbar k. \quad (S26)$$

This is in strong contrast to the behavior of a simple $S = 1/2$ system. The matrices for the $x, y$-components vanish completely. Only the $z$-component has a non-zero matrix,

$$\hat{L}_z + 2\hat{S}_z = \hbar \begin{pmatrix} m & k \\ k & -m \end{pmatrix}. \quad (S27)$$

This entails for the components

$$\frac{\mathrm{d}}{\mathrm{d}t} \langle \hat{L}_x + 2\hat{S}_x \rangle = 0 \quad (S28)$$

$$\frac{\mathrm{d}}{\mathrm{d}t} \langle \hat{L}_y + 2\hat{S}_y \rangle = 0 \quad (S29)$$

$$\frac{\mathrm{d}}{\mathrm{d}t} \langle \hat{L}_z + 2\hat{S}_z \rangle = -\mathrm{i} \left\langle \left[ \begin{pmatrix} E_+ & 0 \\ 0 & E_- \end{pmatrix}, \begin{pmatrix} m & k \\ k & -m \end{pmatrix} \right] \right\rangle - \frac{1}{T_1} \left( \langle \hat{L}_z + 2\hat{S}_z \rangle - \langle \hat{L}_z + 2\hat{S}_z \rangle_{thermal} \right). \quad (S30)$$

Even without expanding the commutator, this looks rather unusual and implies that there is no precession of the magnetic moment. The reason is that the basis states are not eigenstates of the magnetic moment operator as was the case above for the spin up/down states.

The question is, if the magnetic moment operator is indeed the right quantity to look at as we will later need the time dependent conductivity of the system to obtain the tunneling current, which is the quantity that is observed. Hence, we use a generalized polarization function as was already proposed in an earlier publication for the same system (26), where we consider the two eigenstates $\{|+\rangle, |-\rangle\}$ being states having opposite polarization. We can choose

$$\hat{P}_z = P \begin{pmatrix} 1 & 0 \\ 0 & -1 \end{pmatrix} \implies \widehat{H}_0 = \frac{1}{2} \hbar \omega_\mathrm{L} \begin{pmatrix} 1 & 0 \\ 0 & -1 \end{pmatrix} = \frac{1}{2P} \hbar \omega_\mathrm{L} \hat{P}_z, \quad (S31)$$

which corresponds to the $\hat{S}_z$ operator in case of the $S = 1/2$ system. As a measure of the coherence of the system, we can analogously define $\hat{P}_x$ and $\hat{P}_y$ according to the Pauli matrices for the $S = 1/2$ system,

$$\widehat{\mathbf{P}} = P(\sigma_x, \sigma_y, \sigma_z). \quad (S32)$$

This picture is motivated by the fact that at the end, we need to evaluate the tunneling current, where we assume that electrons from the tip that are polarized in a certain direction will have different conductivities for the two states $\{|+\rangle, |-\rangle\}$. If a rotated spin arrives at the atom, i.e. a superposition of spin up and down states, the resulting conductivity will depend on the superposition of the $\{|+\rangle, |-\rangle\}$ states and thus their $\langle \hat{P}_x \rangle$ and $\langle \hat{P}_y \rangle$ expectation values. Therefore, we decouple the tunneling magnetoresistance from the expectation values of the true magnetic moments and rather use the wave function of the atom by assigning a general polarization $\widehat{\mathbf{P}}$ to it. Alternatively, we could have also taken the vector $\vec{n}$ as derived from the density matrix. For this, we would just need to look at the density matrix alone.

To proceed, we need now to express the perturbation Hamiltonian $\widehat{H}'$ in terms of the polarization $\widehat{\mathbf{P}}$. Knowing the matrix elements of $\widehat{H}' = \frac{\mu_\mathrm{B}}{\hbar} (\widehat{\mathbf{L}} + 2\widehat{\mathbf{S}}) \cdot \mathbf{B}_1$, we express the interaction in the polarization,

$$\hat{H}' = \frac{\mu_B}{P} B_{1z}(m\hat{P}_z + k\hat{P}_x) = \frac{2\mu_B}{\hbar} \begin{pmatrix} \hat{P}_x \\ \hat{P}_y \\ \hat{P}_z \end{pmatrix} \cdot \begin{pmatrix} \frac{\hbar k B_{1z}}{2P} \\ 0 \\ \frac{\hbar m B_{1z}}{2P} \end{pmatrix} = -\gamma \hat{\mathbf{P}} \cdot \widetilde{\mathbf{B}}_1, \quad (S33)$$

where $\widetilde{\mathbf{B}}_1 = B_{1z}(\frac{\hbar k}{2P}, 0, \frac{\hbar m}{2P})$ and $\gamma = -2\frac{\mu_B}{\hbar}$ is the same as above.

We notice a couple of things. Apparently, only the $z$-component of the $\mathbf{B}_1$ field plays a role. Further, the $\mathbf{B}_1$ field couples to the $x$-component of the polarization similarly as for the $S = 1/2$ system above, but also to the $z$-component. In principle, also for the $S = 1/2$ system above we would have to assume an oscillating $z$-component since we have no control over the direction of $\mathbf{B}_1$ and can only rotate the coordinate frame to have no y-component. However, we can neglect the $z$-component of $\mathbf{B}_1$ in this case since it only renormalizes the energy difference between the up/down states by a minor amount since $B_1 \ll B_0$.

We will further need the commutator relations for the polarization,

$$[P_i, P_j] = P^2[\sigma_i, \sigma_j] = 2iP^2 \sum_{k=1}^{3} \epsilon_{ijk} \sigma_k = 2iP \sum_{k=1}^{3} \epsilon_{ijk} P_k. \quad (S34)$$

The equation of motion of the polarization function reads then,

$$\frac{d}{dt}\langle\hat{\mathbf{P}}\rangle = -\frac{i}{\hbar}\langle[\hat{H}_0, \hat{\mathbf{P}}]\rangle - \frac{1}{T_1}\left(\langle\hat{\mathbf{P}}\rangle_{\text{longitudinal}} - \langle\hat{\mathbf{P}}\rangle_{\text{thermal}}\right) - \frac{1}{T_2}\langle\hat{\mathbf{P}}\rangle_{\text{decoherence}} \quad (S35)$$

$$= -i\frac{\omega_L}{2P}\langle[\hat{P}_z, \hat{\mathbf{P}}]\rangle - \frac{1}{T_1}\left(\langle\hat{\mathbf{P}}\rangle_{\text{longitudinal}} - \langle\hat{\mathbf{P}}\rangle_{\text{thermal}}\right) - \frac{1}{T_2}\langle\hat{\mathbf{P}}\rangle_{\text{decoherence}}. \quad (S36)$$

The individual components are

$$\frac{d}{dt}\langle\hat{P}_x\rangle = \omega_L\langle\hat{P}_y\rangle - \frac{1}{T_2}\langle\hat{P}_x\rangle \quad (S37)$$

$$\frac{d}{dt}\langle\hat{P}_y\rangle = -\omega_L\langle\hat{P}_x\rangle - \frac{1}{T_2}\langle\hat{P}_y\rangle \quad (S38)$$

$$\frac{d}{dt}\langle\hat{P}_z\rangle = -\frac{1}{T_1}\left(\langle\hat{P}_z\rangle - \langle\hat{P}_z\rangle_{\text{thermal}}\right). \quad (S39)$$

This is the same as we obtained for the spin operators for the spin ½ system. Now, we must add the term $-\frac{i}{\hbar}\langle[\hat{H}', \hat{\mathbf{P}}]\rangle$,

$$-\frac{i}{\hbar}\langle[\hat{H}', \hat{\mathbf{P}}]\rangle = i\frac{\gamma}{\hbar}\langle[\hat{\mathbf{P}} \cdot \widetilde{\mathbf{B}}_1, \hat{\mathbf{P}}]\rangle = -\frac{\gamma 2P}{\hbar}\langle\hat{\mathbf{P}}\rangle \times \widetilde{\mathbf{B}}_1(t) = -\gamma\langle\hat{\mathbf{P}}\rangle \times \begin{pmatrix} k\,B_{1z} \\ 0 \\ m\,B_{1z} \end{pmatrix}. \quad (S40)$$

We will neglect the z-component of $\widetilde{\mathbf{B}}_1$ since it only slightly affects the energy splitting between the $\{|+\rangle, |-\rangle\}$ states and write its $x$-component again as a right and left rotating magnetic field. Putting this altogether we arrive at

$$\frac{d}{dt}\langle \hat{P}_x\rangle = \omega_L \langle \hat{P}_y\rangle - \frac{\omega_1 k}{2}\sin\omega t \langle \hat{P}_z\rangle - \frac{1}{T_2}\langle \hat{P}_x\rangle \quad (S41)$$

$$\frac{d}{dt}\langle \hat{P}_y\rangle = -\omega_L \langle \hat{P}_x\rangle + \frac{\omega_1 k}{2}\cos\omega t \langle \hat{P}_z\rangle - \frac{1}{T_2}\langle \hat{P}_y\rangle \quad (S42)$$

$$\frac{d}{dt}\langle \hat{P}_z\rangle = \frac{\omega_1 k}{2}\left(\sin\omega t \langle \hat{P}_x\rangle - \cos\omega t \langle \hat{P}_y\rangle\right) - \frac{1}{T_1}\left(\langle \hat{P}_z\rangle - \langle \hat{P}_z\rangle_{\text{thermal}}\right). \quad (S43)$$

Here we set $\omega_1 = -\gamma B_{1z}$. The form of these equations is exactly the same as for the spin ½ system and the spin operator. However, the Rabi rate $\omega_1$ is again renormalized, not only by the factor 2 but also by the expectation value k, i.e. the matrix element of the magnetic moment between the two states $\{|+\rangle, |-\rangle\}$. Consequently, the Rabi rate is $\Omega = \frac{\omega_1 k}{2} = \frac{\mu_B}{\hbar} k B_{1z}$.

Altogether, the Bloch-equations as we know them and use them remain the same. The meaning of the expectation values of the spin moment have changed to represent rather the polarization of the system in terms of the ground and first excited state. Also, the origin of the Rabi rate and its strength have been modified, which is the main outcome of this analysis.

**Derivation of the change in tunneling current during EPR**

Since in the STM we measure the long-term evolution of the system, we are interested in the steady-state solution for $\hat{\mathbf{P}}$, i.e., in $\langle \hat{\mathbf{P}}\rangle$. From Eqs. S41-S43, we find

$$\langle \hat{P}_x\rangle = -\langle \hat{P}_z\rangle_{\text{thermal}} \frac{\Omega \Delta\omega T_2^2}{1 + \Delta\omega^2 T_2^2 + \Omega^2 T_1 T_2}, \quad (S44)$$

$$\langle \hat{P}_y\rangle = \langle \hat{P}_z\rangle_{\text{thermal}} \frac{\Omega T_2}{1 + \Delta\omega^2 T_2^2 + \Omega^2 T_1 T_2}, \quad (S45)$$

$$\langle \hat{P}_z\rangle = \langle \hat{P}_z\rangle_{\text{thermal}} \frac{1 + \Delta\omega^2 T_2^2}{1 + \Delta\omega^2 T_2^2 + \Omega^2 T_1 T_2}. \quad (S46)$$

According to Ref. (*4*), the spin resonance is detected electrically through the tunneling magnetoresistance (TMR) effect at the tip-atom junction. Since the STM-junction conductance depends on the relative alignment of the tip spin and the dynamics of the EPR species on the surface, the excited EPR dynamics induces a change of dc as well as ac conductance. While the dc magnetoresistance arising from the time-average population change of surface atom's states is detected by the dc tunneling current, an oscillating conductance at the same frequency as the rf voltage generates a dc tunneling current through mixing with the rf bias, which is called homodyning.

In the EPR-STM measurements, we also need to consider the tip spin $\mathbf{S}^{\text{tip}}$, which is considered as a classical magnetization vector due to the frequent interaction with electrons in the metallic tip that lead to lifetimes in the femtoseconds range (*40*). In the rotating frame, the $z$-component $S_z^{\text{tip}}$ is static. However, the tip magnetization will have a time dependent component in the $xy$-plane given by

$$S_{xy}^{\text{tip}}[\cos(\omega_{\text{rf}}t)\,\boldsymbol{x} + \sin(\omega_{\text{rf}}t)\,\boldsymbol{y}]. \quad (S47)$$

Motivated by classical analogues as in giant-magnetoresistance experiments (*34*), we describe the STM-junction conductance as

$$G = G_J\left(1 + a^{\text{TMR}} \frac{\boldsymbol{S}^{\text{tip}} \cdot \langle \widehat{\boldsymbol{P}} \rangle}{|\boldsymbol{S}^{\text{tip}}||\langle \widehat{\boldsymbol{P}} \rangle|}\right), \quad (S48)$$

where $a^{\text{TMR}}$ is the TMR ratio, i.e., it describes the difference in conductance for the parallel and antiparallel alignment of the respective STM-junction constituents, and $G_J$ is the spin-averaged conductance. We note that Eq. S48 effectively only considers the population of the states $\{|+\rangle, |-\rangle\}$. Future studies might aim at refining Eq. S48 to account also for the off-diagonal components of $\widehat{\boldsymbol{P}}$, i.e., for the coherences, as outlined in Ref. (*26*).

In the experiment, the voltage in the STM junction is given by

$$V = V_{\text{dc}} + V_{\text{rf}} \cos(\omega_{\text{rf}}t + \varphi), \quad (S49)$$

where $\varphi$ accounts for the phase difference between the applied rf voltage and the precession of the EPR species in the lab frame.

The resulting tunneling current reads

$$I(t) = GV = G_J\left(1 + a^{\text{TMR}} \frac{\boldsymbol{S}^{\text{tip}} \cdot \langle \widehat{\boldsymbol{P}} \rangle}{|\boldsymbol{S}^{\text{tip}}||\langle \widehat{\boldsymbol{P}} \rangle|}\right)(V_{\text{dc}} + V_{\text{rf}} \cos(\omega_{\text{rf}}t + \varphi)). \quad (S50)$$

In the experiment, only the dc component of $I(t)$ is detected, which is given by

$$\overline{I(t)} = GV = G_J V_{\text{dc}}\left(1 + a^{\text{TMR}} \frac{S_z^{\text{tip}}\langle \widehat{P_z}\rangle}{|\boldsymbol{S}^{\text{tip}}||\langle \widehat{\boldsymbol{P}} \rangle|}\right) + \frac{1}{2} G_J V_{\text{rf}} \frac{a^{\text{TMR}}}{|\boldsymbol{S}^{\text{tip}}||\langle \widehat{\boldsymbol{P}} \rangle|} S_{xy}^{\text{tip}}(\langle \widehat{P_x}\rangle \cos\varphi - \langle \widehat{P_y}\rangle \sin\varphi). \quad (S51)$$

Using a lock-in amplifier, we measure the difference in $\overline{I(t)}$ between $V_{\text{rf}}$ on and off, i.e.,

$$\Delta I = G_J V_{\text{dc}} \frac{a^{\text{TMR}}}{|\boldsymbol{S}^{\text{tip}}||\langle \widehat{\boldsymbol{P}} \rangle|} S_z^{\text{tip}}(\langle \widehat{P_z}\rangle - \langle \widehat{P_z}\rangle_{\text{thermal}}) + \frac{1}{2} G_J V_{\text{rf}} \frac{a^{\text{TMR}}}{|\boldsymbol{S}^{\text{tip}}||\langle \widehat{\boldsymbol{P}} \rangle|} S_{xy}^{\text{tip}}(\langle \widehat{P_x}\rangle \cos\varphi - \langle \widehat{P_y}\rangle \sin\varphi)$$

$$= -G_J \frac{a^{\text{TMR}}}{|\boldsymbol{S}^{\text{tip}}||\langle \widehat{\boldsymbol{P}} \rangle|} \langle \widehat{P_z}\rangle_{\text{thermal}} \frac{\Omega T_2}{1 + \Delta\omega^2 T_2^2 + \Omega^2 T_1 T_2}\left[S_z^{\text{tip}} V_{\text{dc}} \Omega T_1 \right.$$

$$\left. + \frac{S_{xy}^{\text{tip}}}{2} V_{\text{rf}}(\Delta\omega T_2 \cos\varphi + \sin\varphi)\right]. \quad (S52)$$

By introducing the angle $\alpha$ between the z-axis and the tip spin, and using $G_J \approx I_{\text{dc}}/V_{\text{dc}}$, we find

$$\Delta I = I_{\text{off}} - a^{\text{TMR}} \frac{\langle \widehat{P_z}\rangle_{\text{thermal}}}{|\langle \widehat{\boldsymbol{P}} \rangle|} \frac{\Omega^2 T_1 T_2}{1 + \Delta\omega^2 T_2^2 + \Omega^2 T_1 T_2} I_{\text{dc}}\left[\cos\alpha \right.$$

$$\left. + \sin\alpha \frac{V_{\text{rf}}}{2\Omega T_1 V_{\text{dc}}}(\Delta\omega T_2 \cos\varphi + \sin\varphi)\right], \quad (S53)$$

where an offset current $I_{\text{off}}$ is added to the equation to account for the part of the rf current that is rectified by any nonlinearity of the STM-junction conductance.

Note that $\langle \hat{P}_z \rangle_{\text{thermal}} / |\langle \hat{\mathbf{P}} \rangle|_{\text{thermal}} = 1$, i.e., $\langle \hat{P}_x \rangle_{\text{thermal}} = \langle \hat{P}_y \rangle_{\text{thermal}} = 0$. Additionally, we neglect phase shifts between the rf excitation and the reacting spin in Eq. S53, unlike proposed in Ref. (*4*), since none of the proposed EPR-STM mechanisms involves a resonant process in the GHz range in relating the rf excitation to the EPR driving force. This finally yields Eq. 1 of the main text:

$$\Delta I = I_{\text{off}} - a^{\text{TMR}} I_{\text{dc}} \frac{\Omega^2 T_1 T_2}{1 + \Delta\omega^2 T_2^2 + \Omega^2 T_1 T_2} \left( \cos\alpha + \sin\alpha \frac{\Delta\omega V_{\text{rf}}}{2\Omega T_1 V_{\text{dc}}} \right). \quad (S54)$$

According to Eq. S54, the EPR line shape consists of symmetric and asymmetric parts with respect to $\Delta\omega$. The dc detection of the EPR gives rise to a purely symmetric line shape, whereas the homodyne detection gives rise to an asymmetric line shape. Moreover, we note that Eq. S54 shows that the EPR signal from the homodyne-detection signal does not saturate but scales linearly with $V_{\text{rf}}$ contrary to the dc-detection signal.

**Discussion of the Fano line shape**

According to Ref. (*4*), Eq. S54 can be expressed as a Fano line shape given by

$$I_{\text{Fano}} = I_{\text{off}} + I_{\text{p}} \frac{(q + \delta)^2}{1 + \delta^2}. \quad (S55)$$

However, we stress that fitting the EPR data with such a Fano line shape does not allow for extracting all the physical meaningful parameters given in Eq. S54, since the three parameters defining the Fano line shape, i.e., the amplitude $I_{\text{p}}$, the asymmetry $q$ and the width $\delta$, would depend on each other.

**Discussion of the number of fit parameters**

The model described above (Eq. S54) is derived from a description of the adatom spin by the Bloch equation and the TMR effect in tunneling. The Bloch equations have a minimum number of parameters $(\Omega, T_1, T_2)$ of which we reduced the number of free parameters by connecting the relaxation times to the tunneling current via the parameters $r_1$ and $r_2$, whose values are fixed for each element (making them independent of the experimental conditions). In the TMR model of the tunneling current, only the TMR ratio $a^{\text{TMR}}$ has a fixed value for each element and orientation of the tip magnetization (making it independent again from the experimental conditions for the same STM microtip). In comparison with the model of Ref. (*4*) we refrained from introducing a phase shift in the homodyne tunneling current. All of these assumptions constitute approximations, which allow us to reduce the number of free parameters in the fit.

The free parameters in our model can be summarized in the following way:

One parameter is the same for all spectra: the angle α between the out-of-plane direction and the tip magnetization.

Three parameters are the same for all spectra of a given EPR species: the TMR ratio $a$ and the probability for a tunneling electron to induce an energy ($r_1$) or a phase relaxation event ($r_2$).

Three parameters are different for every spectrum: external field at which the resonance occurs $B_{ext}^0$, the background offset current $I_{off}$ and the Rabi rate $\Omega$.

Thus, in total there are five parameters that have a direct impact on the lineshape in our model (α, $a$, $r_1$, $r_2$ and $\Omega$). Importantly, on average only 1+1/119+3/60≈1.06 parameters are used per spectrum to fit the line shape.

In previous studies, a Fano line shape was used to fit the EPR spectra (4). Such a Fano fit has 3 parameters that determine the line shape (see Eq. S55; $I_p$, $q$, $\delta$). However, it is important to note that the three parameters in this case are free for every single spectrum and a mapping to our (physically more relevant) model is impossible as we also explain in the subsection "Discussion of Fano line shape" above, since the parameters share a common physical base. Thus, for the 119 EPR spectra in our study, one would end up with 119*3=357 free parameters. In stark contrast, our model only employs 119*1+2*3+1=126 free parameters (only $\Omega$ is free for every spectrum, which makes up for 119 of the 126 free parameters; $a$, $r_1$, $r_2$ can change for Fe and TiH). In conclusion, our fitting approach reduces the number of free parameters by a factor of three and additionally provides direct insights into the physically relevant parameters.

## SECTION S3. DETAILED EPR DATA SET

**Full set of 2D plots of EPR data set**

Figure S4 shows the entire EPR data set containing 119 spectra with different $I_{dc}$, $V_{dc}$ and $V_{rf}$ along with fits based on Eq. 1. For TiH, the spectrum for $I_{dc} = 120$ pA and $V_{dc} = 160$ mV is missing because the microtip changed before recording this point.

## SECTION S4. ADDITIONAL FIT RESULTS

Figure S6 A and B shows the asymmetry defined through $T_2 V_{rf}/(2\Omega T_1 V_{dc})$ obtained from the fits of the entire 119 EPR spectra. Note the different scales of the time axis for Fe and TiH. For Fe, the asymmetry is zero within our accuracy.

## SECTION S5. FIT OF THE CALCULATED DISPLACEMENT AND LINEARIZATION IN THE DRIVING ELECTRIC FIELD

We fit the calculated displacements with a second-order polynomial:

$$\Delta z = aE + bE^2, \quad (S56)$$

where $E$ is the electric field. For Fe this gives $a = 2.9 \cdot 10^{-22}$ m²/V and $b = 4.9 \cdot 10^{-32}$ m³/V² and, for Ti, this gives $a = -6.9 \cdot 10^{-22}$ m²/V and $b = -5.1 \cdot 10^{-32}$ m³/V². $E$ is given by the sum of dc and ac electric fields which we obtain by a simple plate capacitor model from the corresponding voltages, which yields:

$$E = E_{\text{dc}} + E_{\text{rf}} \cos \omega_{\text{rf}} t = \frac{V_{\text{dc}} + V_{\text{rf}} \cos \omega_{\text{rf}} t}{s}. \quad (S57)$$

Combining Eqs. (S56)-(S57) yields three displacement terms:

$$\Delta z_0 = aE_{\text{dc}} + bE_{\text{dc}}^2 \quad (S58)$$
$$\Delta z_1 = aE_{\text{rf}} \cos \omega_{\text{rf}} t + 2bE_{\text{dc}}E_{\text{rf}} \cos \omega_{\text{rf}} t \quad (S59)$$
$$\Delta z_2 = \frac{b}{2} E_{\text{rf}}^2 \cos 2\omega_{\text{rf}} t. \quad (S60)$$

The constant displacement $\Delta z_0$ is independent of $E_{\text{rf}}$ and is compensated by the piezo feedback of the STM. The linear term $\Delta z_1$ is proportional to $E_{\text{rf}} \cos \omega_{\text{rf}} t$ and drives the EPR transitions. The second-order term $\Delta z_2$ enables driving of EPR by the second harmonic $2\omega_{\text{rf}}$.

## SECTION S6. DETAILS ON CHARGE-TRANSFER MULTIPLET CALCULATIONS

The model and wave functions that are used to calculate the matrix elements of the magnetic momentum operator $\hat{\boldsymbol{\mu}} = -\mu_B(\hat{\boldsymbol{L}} + 2\hat{\boldsymbol{S}})/\hbar$ between the different magnetic levels of the Fe atom are based on the x-ray absorption spectroscopy (XAS) simulations for the same system as reported in Ref. (*27*). Here, we are only interested in the ground state properties of the Fe atom. The electronic states of the Fe atoms are described using an atomic multiplet model that considers the electron-electron interaction among the valence *d*-electrons using rescaled Slater-Condon integrals for the radial part of the wave functions, and the atomic spin-orbit interaction (*41*). The spherical part of the d-electrons is represented by spherical harmonics $Y_l^m$ with $l = 2$. The atomic environment is simulated by the crystal field potential generated by the surrounding bonding atoms. The finite overlap of the metal wave functions with the ligand atoms (covalency) as well as charge fluctuations in the initial and final states are described by extending the atomic multiplet model to configurational interaction. In such a scheme, in addition to the correlated state of the central atom one considers an additional (delocalized) state or band outside the atom that is generally localized on the ligands (*41-43*). For Fe, this entails a configurational mixing of $d^6$ and $d^7L$ configurations where the $L$ denotes an empty state on the ligand shell (see below). The coupling of this state to the central atom is enabled via a hopping term that effectively annihilates an electron or hole at the ligand orbital and recreates it at the atom site. Different pathways can be distinguished for the hopping term, i.e., electrons can be allowed to hop only onto specific orbitals within the *d*-shell. Thus, the particular symmetry and overlap with the ligand orbitals can be explicitly taken into account. For the calculations the full spectrum of the LS terms is considered. For instance, for a $3d^6$ configuration of an $Fe^{2+}$ this yields 210 different states derived from Slater determinants. The full Hamiltonian is diagonalized considering all contributions (electron-electron interaction, ligand field, spin-orbit coupling and magnetic field) simultaneously using LAPACK routines written in Fortran. This yields wave functions and energies from which we calculate the matrix elements of the spin and orbital moments between any eigenstates of the system. Our code is free of symmetry restrictions, i.e., external fields can be applied in any possible direction.

The crystal field and hopping parameters as well as the charge-transfer energy are determined by systematically varying their values in increasingly narrow energy intervals, starting from an educated guess of their range to achieve a best fit of the simulated x-ray absorption spectra with experiment (*27*). The so obtained parameters are used to calculate the ground state properties. The Slater-Condon integrals are rescaled to 75% because of the overestimation of the Hartree-Fock value and a further reduction due to chemical bonding. The value of the one-electron spin-orbit coupling constant of Fe is taken to be $\xi = 52$ meV for the $d^6$ configuration and $\xi = 45$ meV for the $d^7$ configuration as calculated from Cowan's code (*44*). The charge-transfer energy between the initial state configurations $\Delta = E(d^7L) - E(d^6)$ is set to 0.5 eV to best fit the x-ray data. Similarly, for the Fe/MgO(100) system, the $\sigma$-type bond to the substrate O atom generates an axial crystal field, which we model using the best-fit parameters $D_s = -0.44$ and $D_t = -0.015$ eV. Considering further the $C_{4v}$ symmetry of the adsorption site, we allowed for the $d_{x^2-y^2}$ orbital to interact weakly with the empty Mg states (backbonding) by including a small cubic field of amplitude $10D_q = -0.13$ eV. According to the DFT results, the occupation of the $d-$shell of the Fe ground state configuration is about 6.5 electrons (*27*). To account for this, we considered charge transfer ($\sigma-$donation) between Fe $d-$state and O $2p-$states via the $d_{z^2}$ orbital, by considering the mixing between $d^6$ and $d^7L$ configurations, where $L$ describes a ligand hole on the O site. The hopping parameter amounts to $t_{d_{z^2}} = 0.85$ eV.

As a result, we obtain the wave functions for various magnetic field directions. Figure 4A shows the evolution of the energy levels for an out-of-plane magnetic field swept from zero up to 7 T. The indicated orbital and magnetic moments ($\mu_B = \hbar = 1$) in the figure are the $z-$projections parallel to the magnetic field (moments calculated at 7 T). Assuming that the interaction term of the atom with any magnetic driving field is given by the Zeeman energy term, i.e. $\widehat{H}' = \mu_B(\widehat{L} + 2\widehat{S}) \cdot \boldsymbol{B}/\hbar$, we need to calculate the matrix elements of the spin and oriental angular momentum operators between the ground $|0\rangle$ and first excited $|1\rangle$ state. The numerical values for different field strength and directions are in Figs. S8 (A-C).

**Matrix elements related to the magnetic-moment operator**

For an out-of-plane field applied along the z-direction the in-plane components of the angular momentum operators are essentially zero within the numerical accuracy. For an in-plane field the x- and y-components of the angular momentum matrix elements start to increase but are still smaller by a factor of $10^6$ compared to the z-component matrix elements. This is in strong contrast to a spin ½ system where the z-components vanish and only the x-y components given non-zero matrix elements. In a slightly canted magnetic field off the plane (82° as in Ref. (*1*)), we observe a mixed behavior (see Figs. S8, B-C), where the z-components of the matrix elements decrease with increasing z-projection of the magnetic field and the x- and y- matrix elements increase somewhat with increasing in-plane field component. In any case, the z-matrix elements dominate. This implies that the time varying z-component of the magnetic field drives the EPR transition as opposed to an ordinary spin ½ system.

**Matrix elements related to the crystal-field operators**

Besides magnetic-field-driven EPR, there have been also other mechanisms proposed. One assumes that the electric field drives oscillations of the atoms which causes modulations of the crystal field acting on the magnetic levels of the adatom. Considering the fourfold symmetry of the adatom site the interacting term has been proposed to be proportional to $\widehat{L}_x^4 + \widehat{L}_y^4$. On the other hand, the crystal field parameters are pre-factors for the electronic operators of the type $Y_l^m(\boldsymbol{r})$, which are spherical harmonics. For the fourfold symmetry on the surface, the relevant operators are $Y_2^0$, $Y_4^0$ and $Z_{44} =$ $(Y_4^4 + Y_4^{-4})/\sqrt{2}$. The matrix elements for those operators are plotted in Figs. S8 (D-E). These values are rather small, and an in-plane magnetic field has no effect on them.

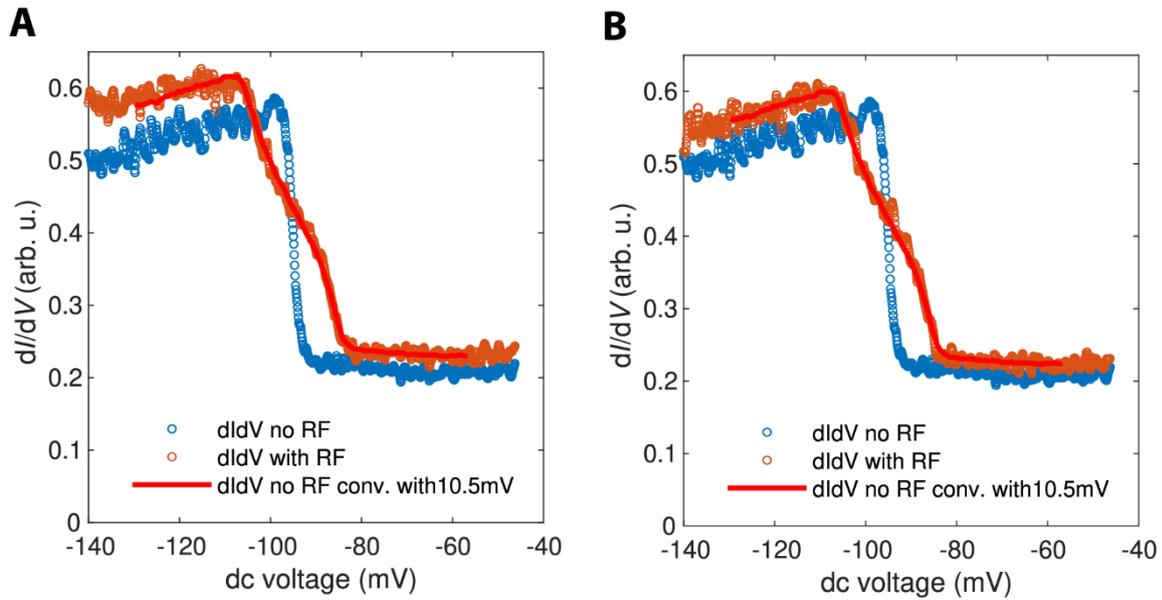

**Figure S1 – Characterization of the rf-voltage amplitude. A** and **B** show the characterization of the rf-transmission function at 8 GHz (A) by the change of the d$I$/d$V$ curve upon application of an rf signal with a power at the rf-signal generator of -20 dBm and at 36 GHz (B) by the change of the d$I$/d$V$ curve upon application of an rf signal with a power at the rf-signal generator of -13.13 dBm. The solid line shows a convolution of the bare d$I$/d$V$ curve (blue circles) with an rf-voltage amplitude of 10.5 mV.

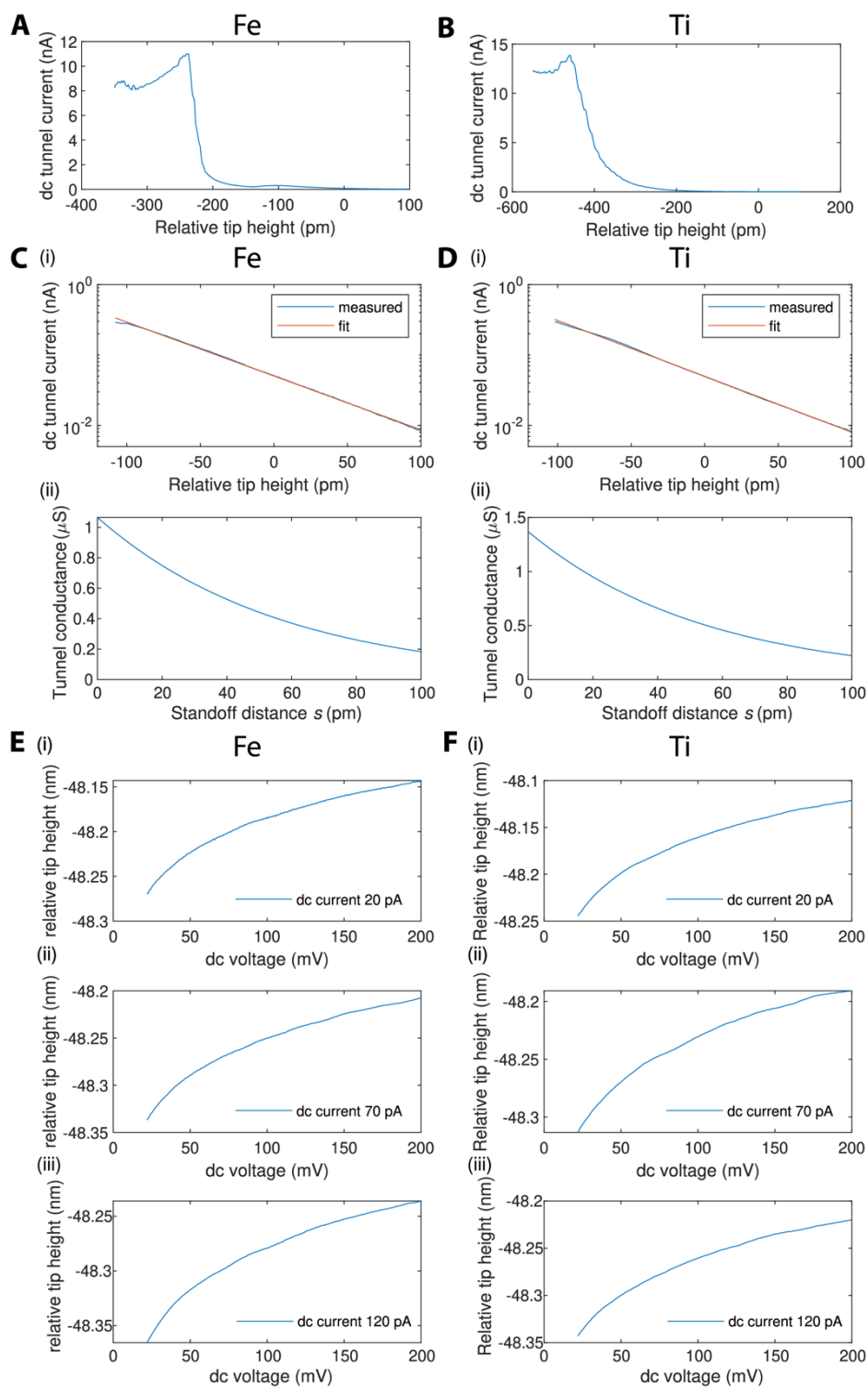

**Figure S2 – Characterization of the standoff distance. A** and **B** show a point-contact measurement on Fe (A) and hydrogenated Ti (B) on the bridge binding site. Feedback opened at a dc bias of 10 mV. **C** and **D** show the absolute height calibration for the Fe (C) and hydrogenated Ti (D) atom used for recording the EPR data set. (i) $I(z)$ curve along with an exponential fit. Feedback opened at $V_{dc} = 10$ mV. (ii) From the exponential fit in (i), and the point-contact conductance (A and B), the absolute standoff distance is determined. **E** and **F** show the relative tip height vs dc voltage on Fe (E) and hydrogenated Ti (F), where the feedback is opened at the three dc currents [(i)-(iii)] used for the EPR sweeps (see Fig. S4).

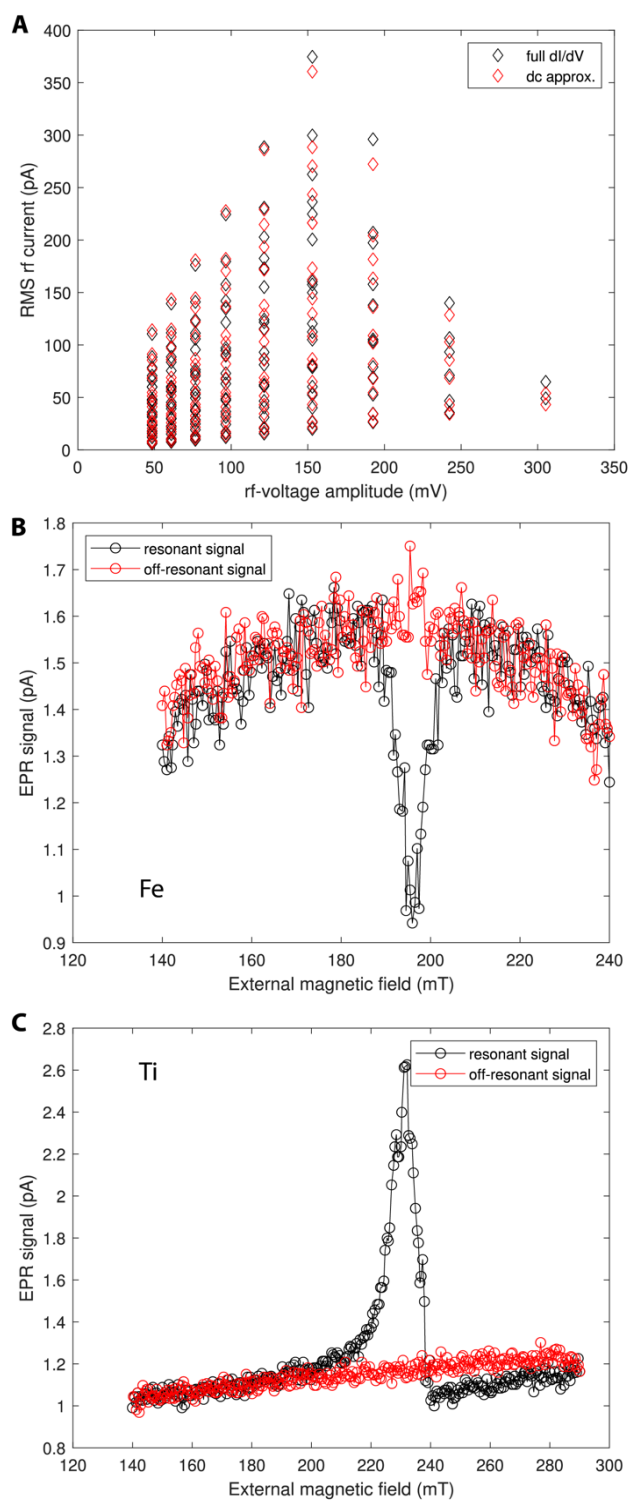

**Figure S3 – Characterization of additional EPR parameters. A** Root-mean-square (RMS) rf current calculated with two distinct approaches. Black symbols: Convoluting the experimental d$I$/d$V$ curve with sinusoidal rf voltage of amplitude $V_{\text{rf}}$ over one period. Red symbols: Calculating directly the rf current via the setpoint dc resistance and Ohm's law. **B** and **C** show exemplary EPR spectrum on Fe (B) recorded at an rf frequency of 36 GHz (black) along with an off-resonant reference sweep at 8 GHz (red) and on hydrogenated Ti (C) recorded at an rf frequency of 8 GHz (black) along with an off-resonant reference sweep at 36 GHz (red).

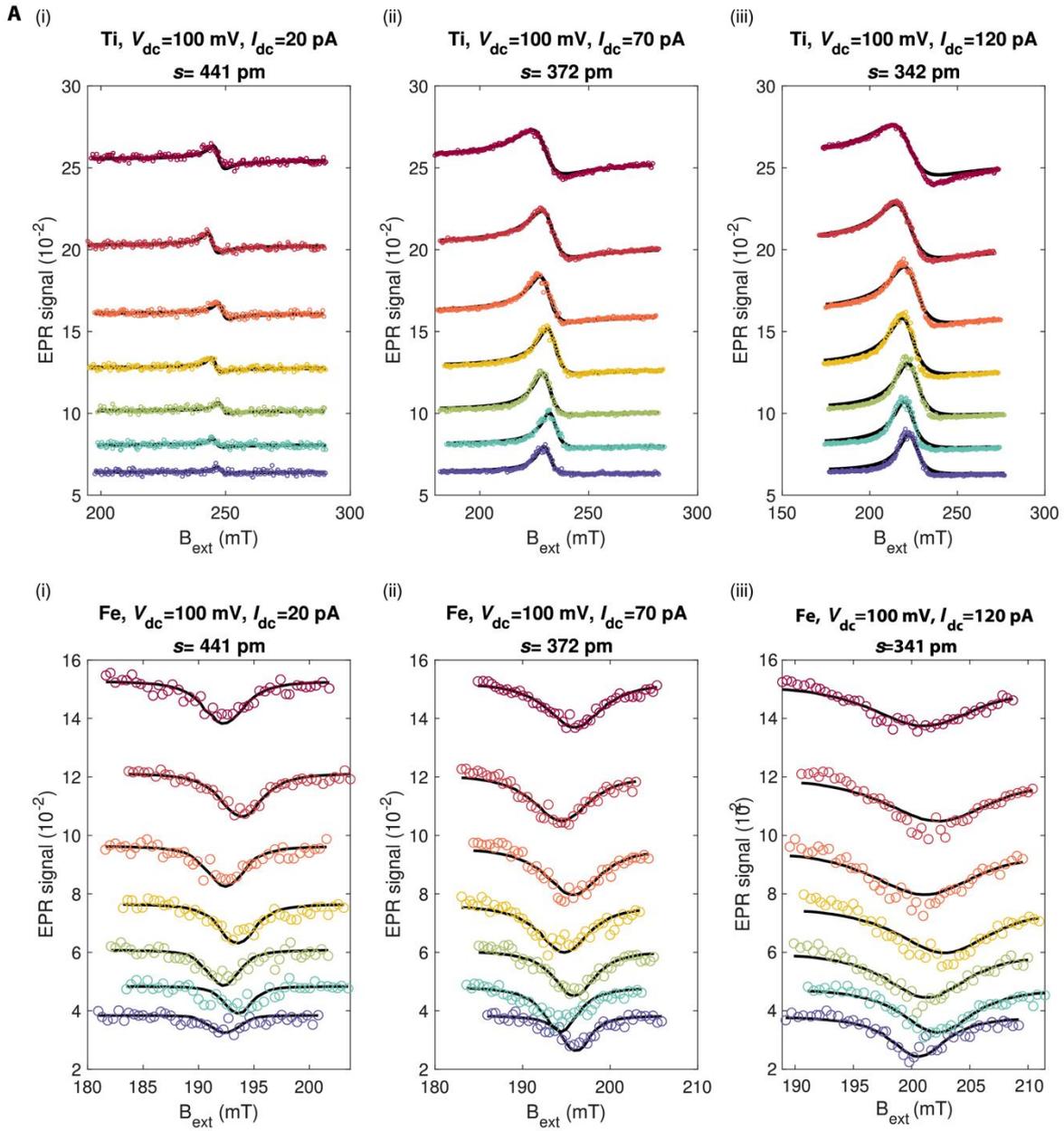

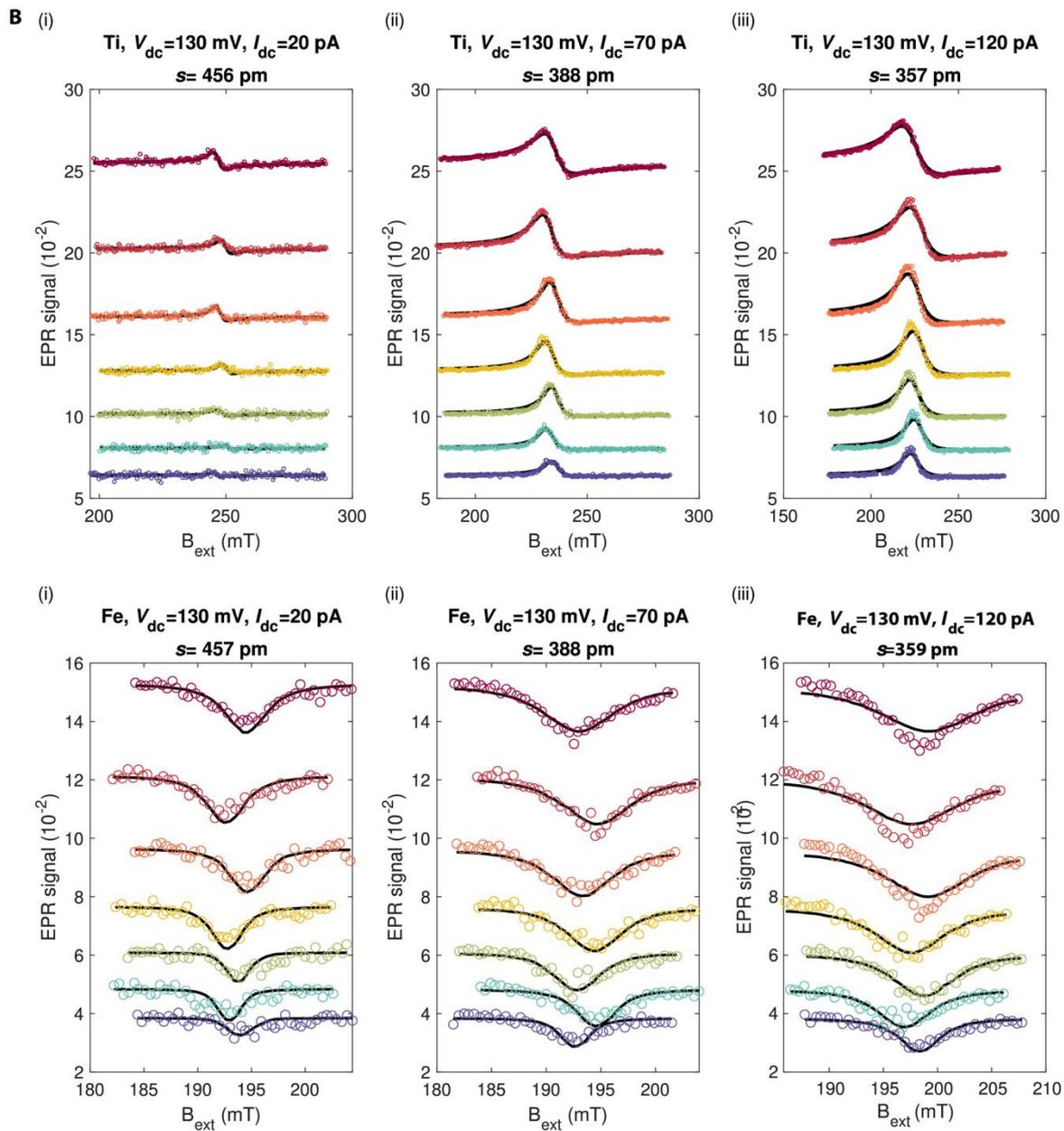

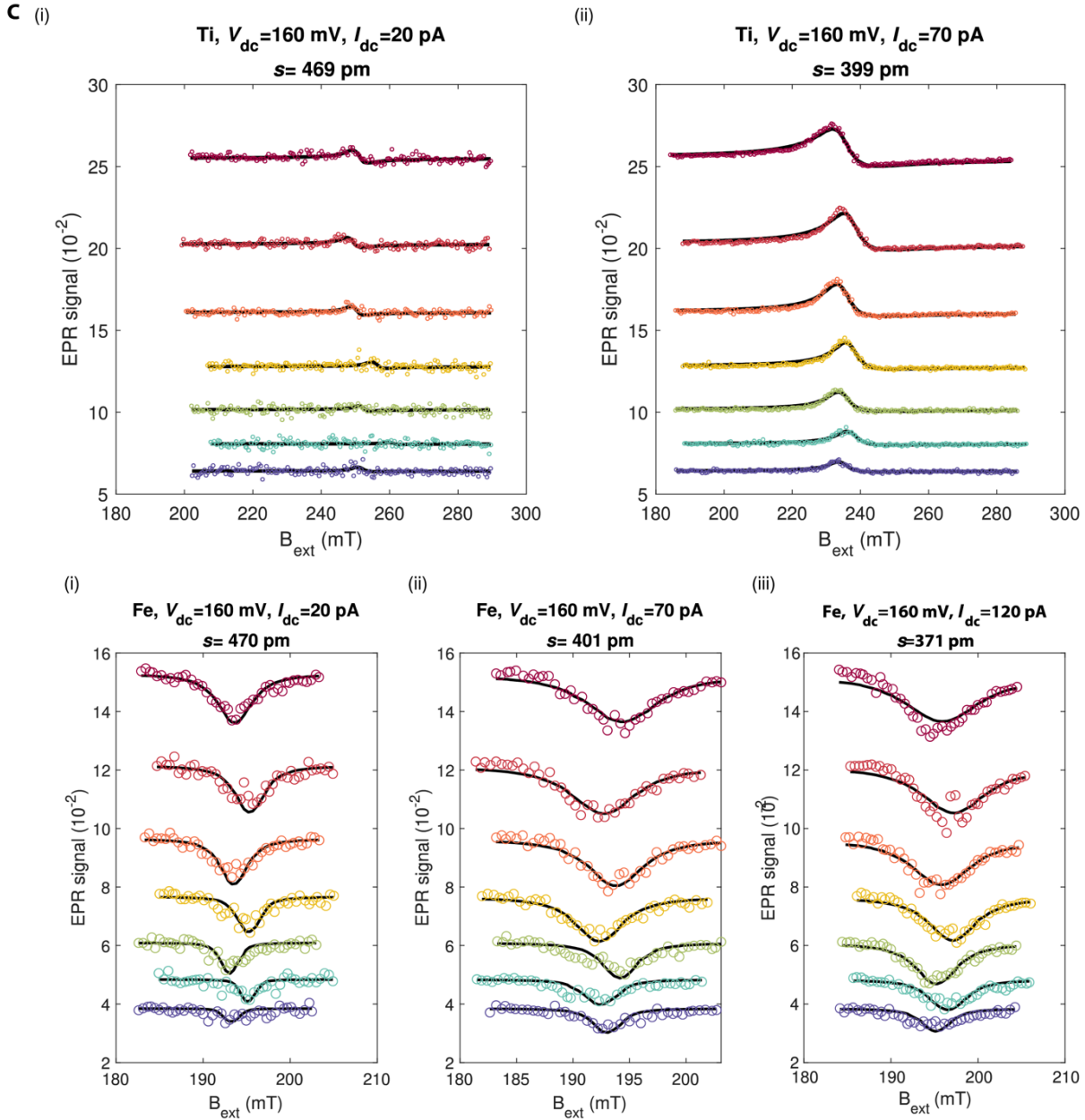

**Figure S4 – Detailed EPR data set with fits.** EPR spectra along with fits (solid lines) based on Eq. (1) for hydrogenated Ti [(a)-(c)] and Fe [(d)-(f)] at a dc bias of $V_{dc} = 100$ mV (**A**), $V_{dc} = 130$ mV (**B**) and $V_{dc} = 160$ mV (**C**) for different dc currents $I_{dc}$ and standoff distances $s$ as indicated in the titles of the subplots [(i)-(iv)]. For better visibility, EPR spectra are background-corrected, normalized by $I_{dc}$, and vertically offset. The vertical offset is proportional to the rf voltage amplitude $V_{rf}$, which is color-coded and in the order from bottom to top: 64 mV, 81 mV, 102 mV, 128 mV, 161 mV, 203 mV and 256 mV.

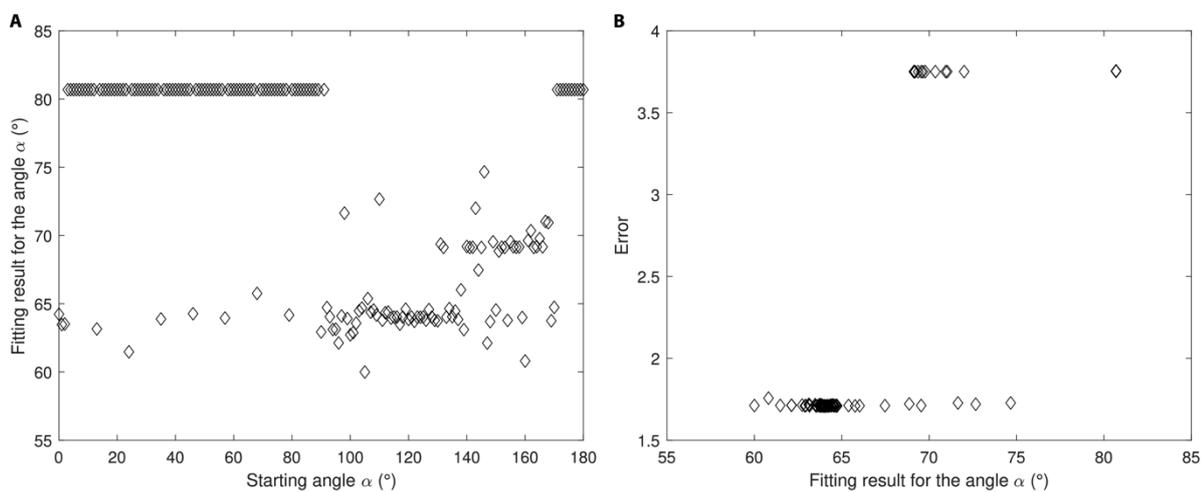

**Figure S5 – Details on fit procedure. A** Fit results for the angle α for varying starting values of α. **B** Error of the fits for the different obtained best-fit values of α. The results with the smallest errors, i.e., the best fits cluster around α = 64°.

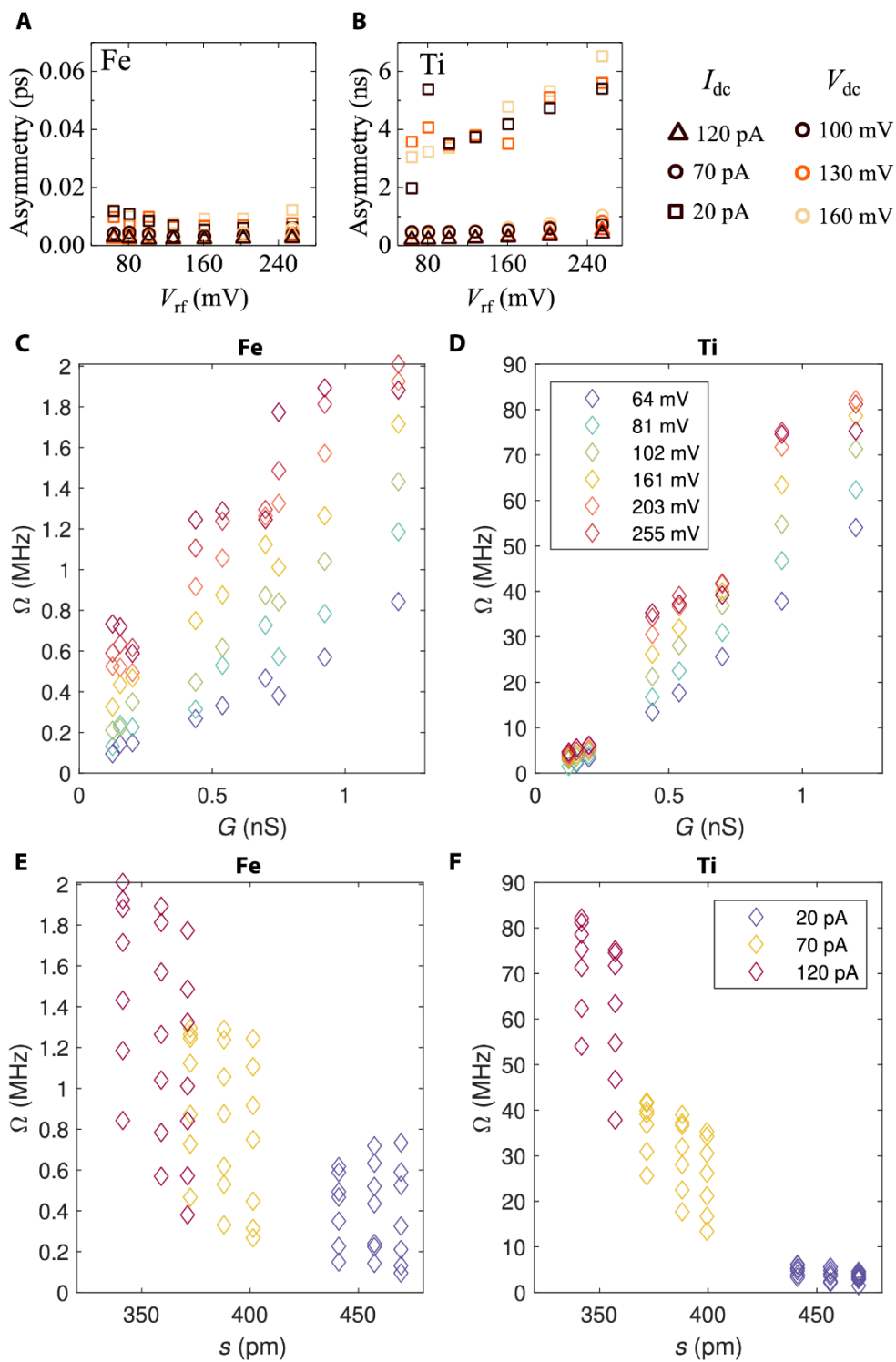

**Figure S6 – Additional EPR-spectra fit results.** Asymmetry extracted from the fits of the entire data set (see main text for details) for Fe (**A**) and hydrogenated Ti (**B**) for varying experimental conditions. Note the ps time scale for Fe. Experimental Rabi Ω rate vs junction conductance $G$ for Fe (**C**) and hydrogenated Ti (**D**). In C and D, colors represent different values of the rf-voltage amplitude. Experimental Rabi Ω rate vs standoff distance $s$ for Fe (**E**) and hydrogenated Ti (**F**). In E and F, colors represent different values of the setpoint dc current.

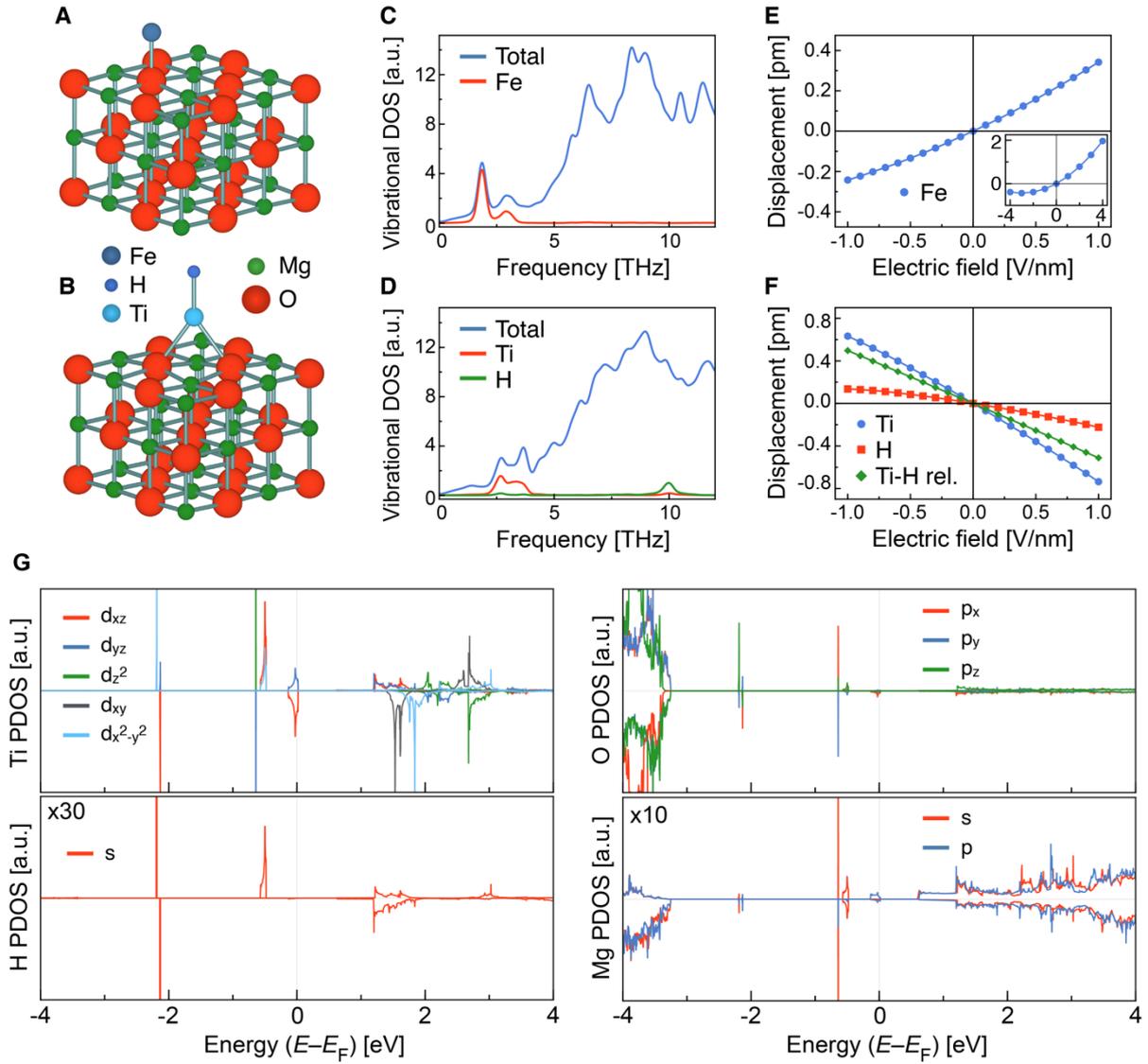

**Figure S7 – Details on DFT calculations.** **A** and **B** show the unit cells of Fe (oxygen binding site) and TiH (bridge binding site), respectively, on three monolayers of MgO used in the calculations. **C** and **D** show the total and partial vibrational density of states (DOS) of the respective systems, revealing the low-frequency localized vibrational modes of the adatoms. **E** and **F** show the displacements of Fe and TiH with respect to the MgO surface, induced by a static electric field between ±1 V/nm applied normal to it. The inset in E shows a larger range of electric fields, in which the asymmetry in displacement for Fe becomes apparent. **G** shows the projected density of states for the TiH species. The spin-dependent densities of states for the elements Ti, H, O and Mg, which are involved in the binding of the TiH species, are decomposed into the most relevant valence orbitals in the vicinity of the Fermi energy.

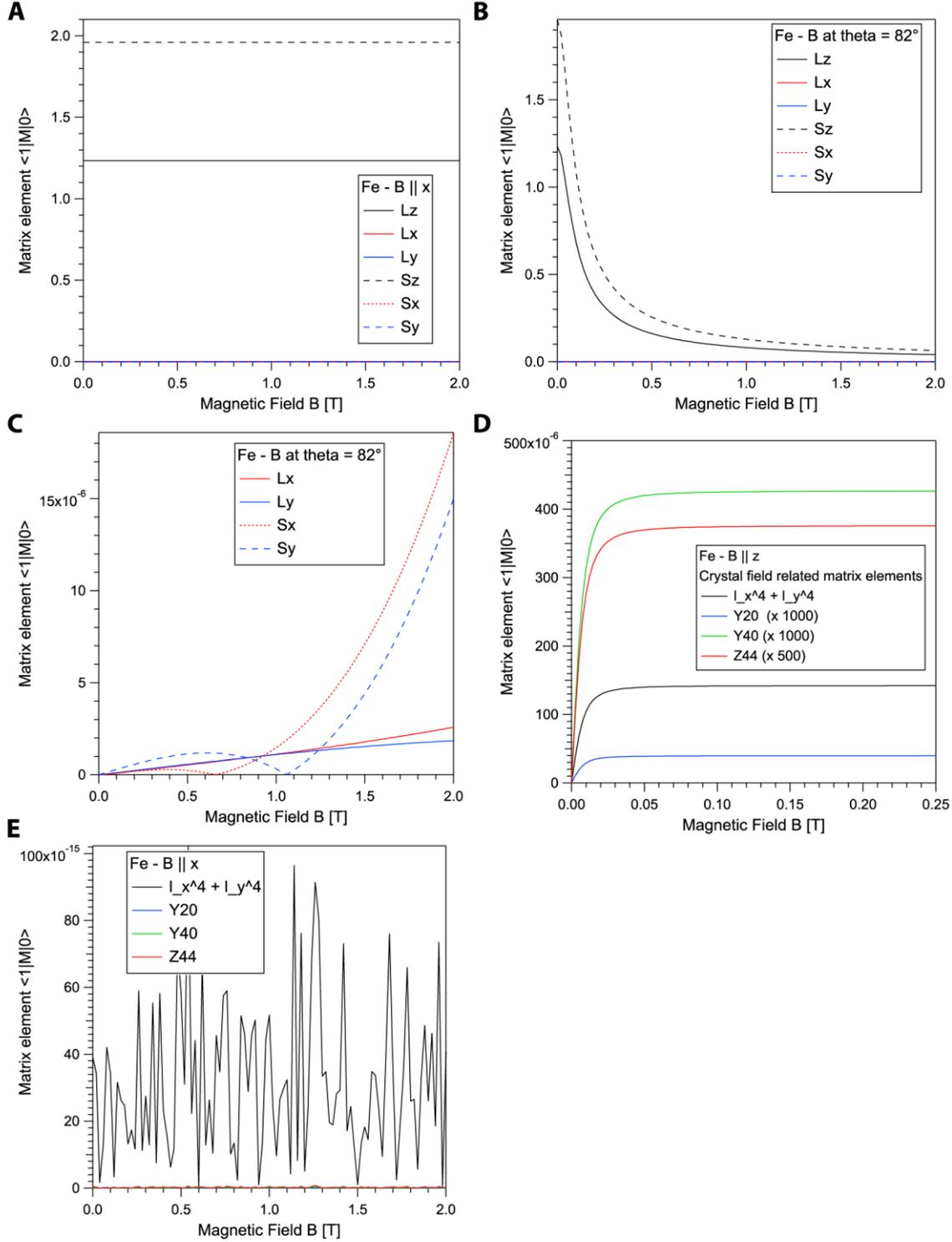

**Fig. S8. Details on multiplet calculations. A** Calculated components of the matrix elements of the orbital and spin momentum operator $\hat{L}$ and $\hat{S}$, respectively, of the Fe/MgO/Ag(100) system for an external magnetic field perpendicular to $z$. **B** shows calculated components of the matrix elements of the orbital and spin momentum operator $\hat{L}$ and $\hat{S}$, respectively, of the Fe/MgO/Ag(100) system for an external magnetic field canted with respect to the z-axis by 82°. **C** shows a magnified view of B for the in-plane components. **D** and **E** show calculated matrix elements for Fe/MgO/Ag(100) for the crystal-field operators for an external magnetic field along $z$ (D) and perpendicular to $z$ (E).

**A**

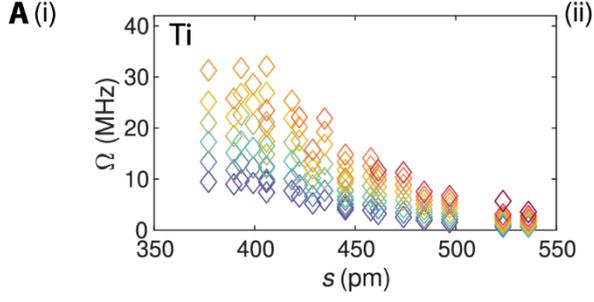
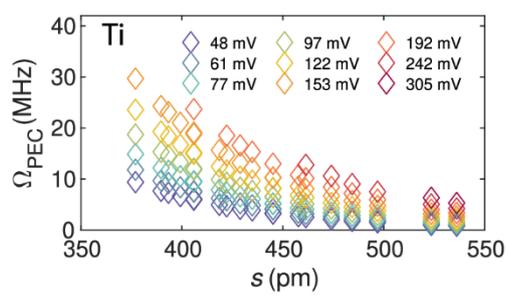

| # TiH$_B$ spectra | 135 |
|---|---|
| $I_{dc}$ (pA) | 20, 40, 60, 80, 120, 160, 200 |
| $V_{dc}$ (mV) | 60, 80, 100 |
| $V_{rf}$ (mV) | 48, 61, 77, 97, 122, 153, 192, 242, 305 |

| α | 19° |
|---|---|
| $r_1^{Ti}$ | 0.1 |
| $r_2^{Ti}$ | 1.0 |
| $a_{TMR}^{Ti}$ | −6.4 |
| $\lambda^{Ti}$ (pm) | 75 |
| $a^{Ti}$ (T) | −40.0 |
| $b$ ($\mu_0\mu_B$) | 0.1 |
| add. broadening (mT) | 1 |

**B**

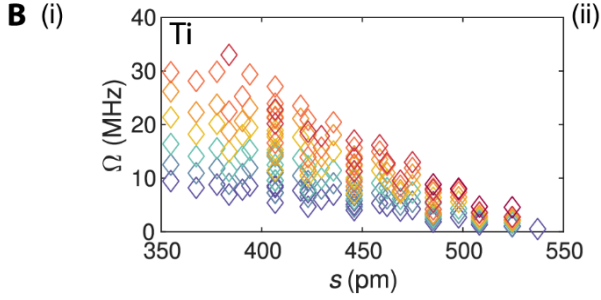
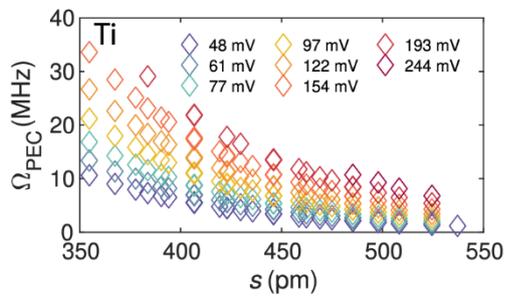

| # TiH$_B$ spectra | 184 |
|---|---|
| $I_{dc}$ (pA) | 20, 40, 60, 80, 120, 160, 200 |
| $V_{dc}$ (mV) | 40, 60, 80, 100 |
| $V_{rf}$ (mV) | 48, 61, 77, 97, 122, 154, 193, 244 |

| α | 16° |
|---|---|
| $r_1^{Ti}$ | 0.1 |
| $r_2^{Ti}$ | 1.0 |
| $a_{TMR}^{Ti}$ | −6.6 |
| $\lambda^{Ti}$ (pm) | 100 |
| $a^{Ti}$ (T) | −12.8 |
| $b$ ($\mu_0\mu_B$) | 0.4 |
| add. broadening (mT) | 1 |

**C**

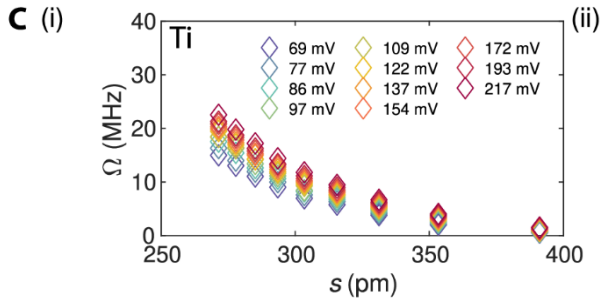
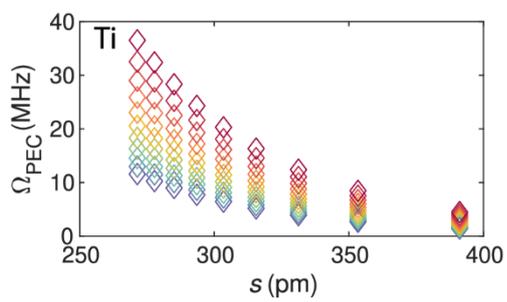

| # TiH$_B$ spectra | 99 |
|---|---|
| $I_{dc}$ (pA) | 20, 40, 60, 80, 100, 120, 160, 180 |
| $V_{dc}$ (mV) | 100 |
| $V_{rf}$ (mV) | 69, 77, 86, 97, 109, 122, 137, 154, 172, 193, 217, |

| α | 12° |
|---|---|
| $r_1^{Ti}$ | 0.3 |
| $r_2^{Ti}$ | 0.7 |
| $a_{TMR}^{Ti}$ | −38.3 |
| $\lambda^{Ti}$ (pm) | 70 |
| $a^{Ti}$ (T) | −11.3 |
| $b$ ($\mu_0\mu_B$) | 0.1 |
| add. broadening (mT) | 1 |

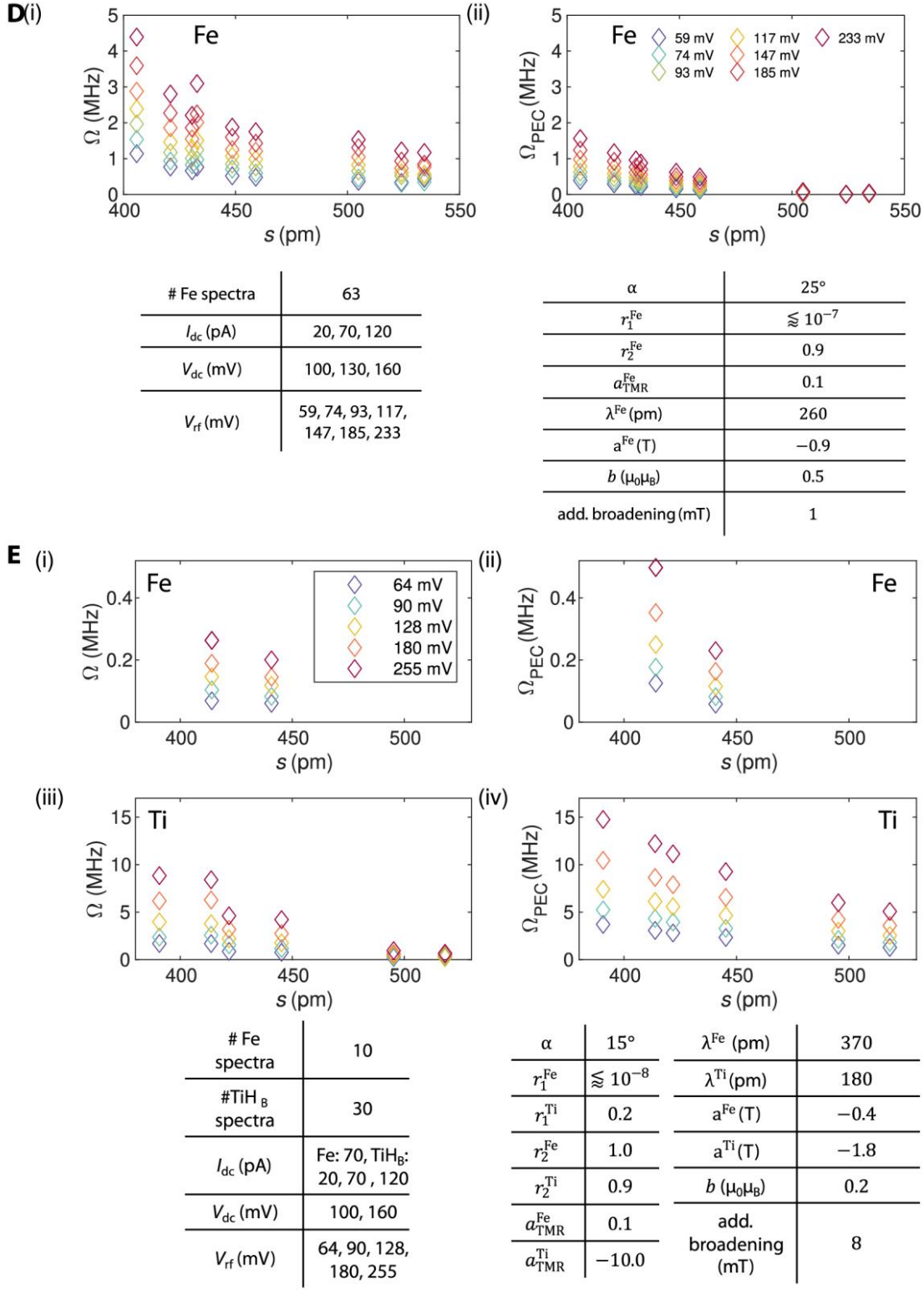

**Figure S9 – Evaluating the Rabi rate for additional large EPR-STM data sets. A − E** show the experimental Rabi rate $\Omega$ in comparison with the calculated Rabi rate $\Omega_{PEC}$ vs the standoff distance $s$ for different microtips. Tables below the respective data indicate the experimental and the fit parameters (compare with main text).

| Parameter | Range |
|---|---|
| $\alpha$ | $[0, \pi/2]$ |
| $a_{\text{TMR}}^{\text{Fe}}$ | $[-1,1]$ |
| $a_{\text{TMR}}^{\text{Ti}}$ | $[-1,1]$ |
| $r_1^{\text{Fe}}$ | $[0,0.5]$ |
| $r_1^{\text{Ti}}$ | $[0,0.5]$ |
| $r_2^{\text{Fe}}$ | $[0.5,1]$ |
| $r_2^{\text{Ti}}$ | $[0.5,1]$ |
| $\Omega$ | $[10^7 \text{ Hz}, 10^{10} \text{ Hz}]$ |
| $b$ | $[0, 5\ \mu_0\mu_B]$ |
| $a^{\text{Fe}}$ | $[-10\text{ T}, 10\text{ T}]$ |
| $a^{\text{Ti}}$ | $[-10\text{ T}, 10\text{ T}]$ |
| $\lambda_{\text{rel}}^{\text{Fe}}$ | $[30\text{ pm}, 500\text{ pm}]$ |
| $\lambda_{\text{rel}}^{\text{Ti}}$ | $[30\text{ pm}, 500\text{ pm}]$ |

**Table S1** – Range of fit parameters for the data set in Figure 2.